\documentclass{JFM-FLM_Au}
\usepackage{graphicx}
\usepackage{amsmath}
\usepackage[table,dvipsnames]{xcolor}
\usepackage{hyperref}
\usepackage[T1]{fontenc}
\usepackage{siunitx}
\usepackage{enumitem}
\usepackage[utf8]{inputenc} 
\newcommand\Stokes{\mbox{\textit{St}}}  
\newcommand\Peclet{\mbox{\textit{Pe}}}  
\let\v\vold
\newcommand{\v}[1]{\boldsymbol{#1}}
\newcommand{\vhat}[1]{\hat{\boldsymbol{#1}}}
\newcommand{\tensor}[1]{\mathsfbi{#1}}

\lefttitle{Pierce et al.}
\righttitle{Journal of Fluid Mechanics}
\title{How pore-scale disorder controls fluid stretching in two-dimensional porous media}
\author{J. Kevin Pierce\aff{1}, Tanguy Le Borgne\aff{1,2}, Fran\c{c}ois Renard\aff{1,3}, \and Gaute Linga\aff{1,4}}
\affiliation{
\aff{1}PoreLab, The Njord Centre, Departments of Physics and Geosciences, University of Oslo, Norway
\aff{2}G\'{e}osciences Rennes, UMR 6118 CNRS University of Rennes, Rennes, France
\aff{3}Univ. Grenoble Alpes, Univ. Savoie Mont Blanc, CNRS, IRD, Univ. Gustave Eiffel, ISTerre, 38000 Grenoble, France
\aff{4}PoreLab, Department of Physics, Norwegian University of Science and Technology, Trondheim, Norway
}
\corresau{K. Pierce, kevinpi@uio.no}
\begin{document}
\maketitle
\begin{abstract}
Fluid stretching in porous media governs the mixing of reactants, contaminants, and nutrients, yet how the solid microstructure controls the stretching statistics remains poorly understood.
Here, we investigate how porous-medium heterogeneity controls fluid stretching at the pore scale using (i) particle-tracking velocimetry experiments in 3D-printed millifluidic cells, (ii) numerical simulations of solute plume stretching in the measured flow fields, and (iii) analytical calculations of fluid elongation.
The cells contain arrays of cylinders arranged with systematically controlled disorder levels, from ordered to random.
Velocity and shear-rate measurements show that fluid stretching is strongly localized near solid boundaries for all disorder levels, indicating that near-wall flow is the main driver of stretching. 
The mean elongation grows linearly in time for ordered media and quadratically for disordered media.
This nonlinear scaling arises from repeated cylinder encounters that progressively increase the number of filaments stretched by shear, with the encounter frequency directly related to the pore-scale heterogeneity.
Elongation distributions have broad tails but deviate from log-normal distributions, consistent with the additive structure of stretching in two-dimensional steady flows.
We analytically describe the stretching produced by flow around an isolated cylinder and embed this description in a random-walk framework that predicts stretching statistics in disordered media.
This framework quantitatively explains the observed elongation moments and provides a new conceptual foundation to understand fluid elongation in heterogeneous media beyond the common representation as a simple shear flow.
\end{abstract}
\section{Introduction}
Solute mixing in porous media flows underlies numerous scientific applications, including contaminant transport and reactions in aquifers \citep{kitanidis_delivery_2012, valocchi_mixinglimited_2019, rolle_mixing_2019}, microbial processes in the subsurface \citep{borer_spatial_2018, turuban_cooperative_2025}, and carbon dioxide sequestration in the subsurface \citep{szulczewski_lifetime_2012, snaebjornsdottir_carbon_2020}.
Mixing describes the process by which initially segregated chemicals diffuse toward uniformity due to the advective stirring of a flow field \citep{ottino_kinematics_2004, villermaux_mixing_2019}.
Porous media flows provide complex velocity fields, which stretch solute fields into elongated filaments and sharpen transverse concentration gradients \citep{leborgne_fluid_2026}.
Because concentration gradients drive diffusion, predicting the evolution of concentration fields in porous media can be an intricate problem, even in non-turbulent flows \citep{dentz_mixing_2023}.
Fluid deformation also couples to diverse phenomena of practical importance in addition to mixing, including the stability of visco-elastic fluids \citep{mokhtariBirefringentStrandsDrive2022, kumarLagrangianStretchingReveals2023}, the transport of colloids and fibers \citep{cunhaSettlingDynamicsBrownian2022,duroureDynamicsFlexibleFibers2019}, and dynamics of immiscible droplets \citep{stoneDYNAMICSDROPDEFORMATION2026}.

The lamellar theory provides a framework for relating mixing rates to flow-induced stretching \citep{ranz_applications_1979, ottino_framework_1981, meunier_how_2003, dimotakis_turbulence_1999, leborgne_lamellar_2015, souzy_mixing_2018, dentz_mixing_2023}.
This framework decomposes solute plumes into infinitesimal strips (the lamellae) and relates the evolution of solute concentrations to the stretching history of the individual lamellae \citep{leborgne_lamellar_2015, villermaux_mixing_2019}.
While fluid stretching has long been linked to the distributions of velocity and shear \citep{ottino_mixing_1989, hinch_mixing_1999, dentz_coupled_2016, lester_fluid_2018}, our understanding of how the interior structure of porous media controls the velocity and shear rate distributions remains limited \citep{anna_mixing_2014, dentz_continuous_2016, dentz_mechanisms_2018, alim_local_2017, ben-noah_pore_2024}.
Although solute mixing and chemical reactions have been quantified in a variety of laboratory and digital porous media \citep{willingham_evaluation_2008, souzy_mixing_2018, heyman_stretching_2020, dentz_evolution_2018, borgman_solute_2023,shafabakhsh_resolving_2024,shafabakhsh_coupling_2025}, there has been relatively little progress in connecting the underlying stretching statistics to the porous structure.
Accordingly, we still lack the capability to explain mixing and reaction rates from the porous medium structure alone \citep{leborgne_fluid_2026}.

The statistics of fluid elongation scale differently with time in two- and three-dimensional steady flows \citep{leborgne_fluid_2026}.
In two-dimensional steady flows, the elongation moments grow algebraically with time ($\langle \rho^q \rangle \sim  t^{\alpha(q)}$), whether in open flows \citep{souzy_mixing_2018, meunier_how_2003} or in porous media \citep{leborgne_lamellar_2015,dentz_coupled_2016, lester_fluid_2018}.
The mean elongation exponent ($q=1$) ranges between $\alpha(1)=0.5$ (sub-linear) and $\alpha(1)=2$ (quadratic) in porous media \citep{dentz_coupled_2016, leborgne_fluid_2026}, with a higher exponent in more heterogeneous media, at least in random-permeability Darcy flows \citep{leborgne_lamellar_2015}.
Earlier studies have assumed that linear mean elongation ($\alpha(1)=1$) applies in disordered media to successfully explain some observed fluid-mixing properties \citep{jimenez-martinez_impact_2017,borgman_solute_2023}.
Understanding the limits of this linear-elongation assumption in two-dimensional porous media and relating the higher elongation moments to the medium structure remain significant open problems.

In three-dimensional porous media, the flow can be chaotic, with elongation moments growing exponentially with time \citep{heyman_stretching_2020, souzy_velocity_2020}, unless additional symmetries are present in the medium structure \citep{turuban_spacegroup_2018,turuban_chaotic_2019}.
The mean elongation is characterized by a Lyapunov exponent, which has been quantitatively linked to the frequency of branch points in the flow \citep{lester_unified_2025}.
However, specific relations between the Lyapunov exponent and pore structure have only been established for monodisperse bead-packs \citep{turuban_chaotic_2019, leborgne_fluid_2026}.
A clearer understanding of the two-dimensional case, where the pore structure can be carefully controlled, may provide insights for the three-dimensional problem.

Beyond the scaling of moments, the full elongation distribution $P(\rho,t)$ is needed to predict solute concentrations from the lamellar theory \citep{leborgne_lamellar_2015}.
Whether in two or three dimensions, the overall elongation distribution is often considered to be log-normal, based on the reasoning that the coarse-grained stretching process is multiplicative with uncorrelated increments \citep{leborgne_lamellar_2015, villermaux_mixing_2019, souzy_velocity_2020}.
This idea has been carried over from the literature on turbulent flows \citep{kraichnan_convection_1974, drummond_turbulent_1990, duplat_nonsequential_2010}.
Whether log-normal stretching is appropriate for two-dimensional steady porous media flows remains to be clarified, especially since the exact description of stretching in two-dimensional steady flows is inherently additive, not multiplicative \citep{dentz_coupled_2016}.
With additive stretching, log-normal elongation distributions are unexpected; yet log-normal-like elongation distributions have been observed in some two-dimensional porous media flows \citep{leborgne_lamellar_2015}.

In the present study, we evaluate how fluid elongation depends on the medium structure in quasi two-dimensional porous models composed of cylinder arrays.
Specifically, we combine particle tracking velocimetry experiments and numerical simulations to link the temporal scaling exponents of the lower elongation moments to the porous medium heterogeneity.
We observe that the mean fluid elongation grows in time linearly along a symmetry direction of an ordered cylinder array, while it grows quadratically in disordered arrays with a growth rate that depends on the medium heterogeneity.
By considering the frequency with which the plume encounters cylinders during transport, we relate the quadratic scaling to the progressive wrapping of the stretched plume over cylinders which increases the number of branches stretched by shear.
To describe our observations, we analytically calculate the stretching produced by flow around an isolated cylinder, and we represent elongation in cylinder arrays as a stochastic exchange of individual lamellae between near-wall and bulk-flow regions.
The isolated cylinder results are used to parameterize the near-wall stretching in the stochastic exchange model for elongation.
The observed and modeled elongation statistics align with the Protean-frame theory of \citet{dentz_coupled_2016}, while also going further to link the elongation statistics observed in porous media to characteristics of the underlying solid structure.
This work provides a framework to link pore-scale disorder to fluid elongation and mixing in steady flows through two-dimensional porous media.

\section{Methods}
\begin{figure}[htb]
    \centering    \includegraphics[width=\linewidth]{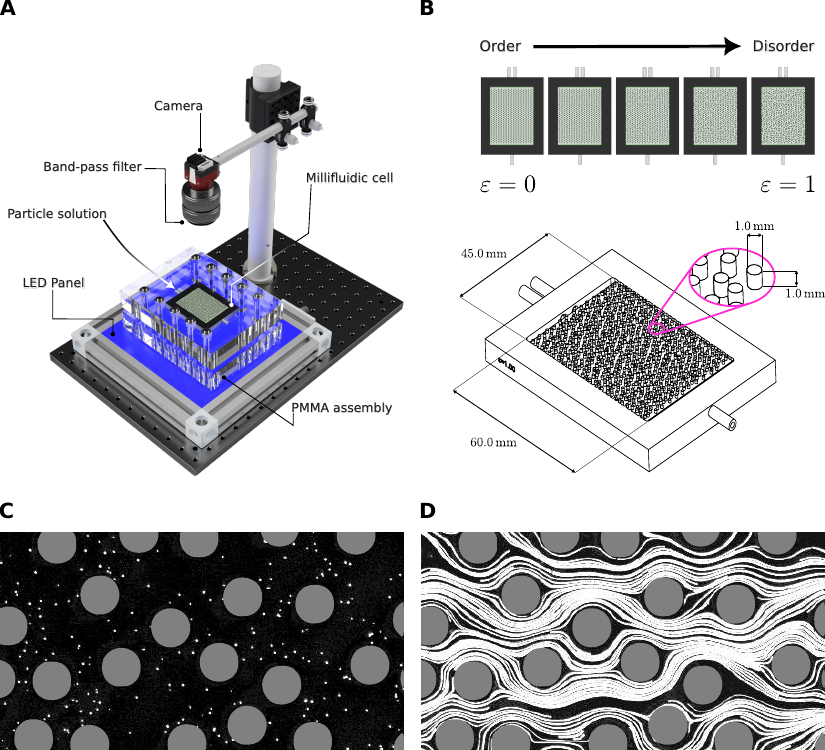}
    \caption{(A) Schematic of the experimental setup for particle-tracking velocimetry. (B) Detail of the progressively-disordered stereolithography-printed cells, with vertical rods of diameter $a$ arranged according to a disorder parameter $\varepsilon$. (C) Zoomed-in view of an experimental image, with fluorescent microspheres visible as white dots. (D) Overlay of successive zoomed-in images, showing particles tracing out streamlines.}
    \label{fig:exptsetup}
\end{figure}
\subsection{Flow cell fabrication and structure}
Transparent millifluidic porous flow cells were fabricated with \SI{20}{\micro\meter} resolution using a stereolithographic 3D-printer (Formlabs 3B) and clear polyacrylate resin (Formlabs Clear V4). Models were designed using Autodesk Fusion CAD software as rectangular cavities with dimensions $60\times45\times\SI{1}{\milli\meter\cubed}$, filled with vertical cylinders of diameter $a=\SI{1}{\milli\meter}$ and average spacing $2 a$ between centers. 
The vertical gap was set to be much smaller than the transverse size of the cell, which enables us to consider the flow as approximately two-dimensional \citep{roman_particle_2016,jasinski_effective_2018}. The Autodesk Fusion Python API was used to place the cylinders at specified locations in the domains. Models were printed with open tops for easy cleaning, and the top surfaces were painted black to contrast the solid. Models were sealed during experiments by pressing them between thick poly(methyl methacrylate) plates (PMMA) using thin polyvinyl chloride (PVC) sheets as gaskets.

Cylinders were placed at random sites $\v{r}'_{ij}$ resulting from perturbations of a triangular lattice. The perturbation scheme is based on a dimensionless parameter $\varepsilon$ that ranges between $0$ and $1$.
Unperturbed sites are given by $\v{r}_{ij} = i \v{a}_1 + j \v{a}_2 $, where $i,j$ are integers and $\v{a}_k$ are the primitive translation vectors of the triangular lattice. Perturbed sites are defined by
\begin{equation}
   \v{r}'_{ij} = \v{r}_{ij} + \frac{a\varepsilon}{2}\sqrt{n_{ij}}\left(\cos(\theta_{ij})\vhat{e}_1 +  \sin(\theta_{ij})\vhat{e}_2\right),
\end{equation}
where $0 \leq n_{ij}\leq 1$ and $0\leq \theta_{ij} \leq 2\pi $ are uniform random variables and the $\vhat{e}_i$ are unit vectors parallel and perpendicular to the mean flow direction. This specification places the cylinders uniformly within disks of diameter $a(1+\varepsilon)$ centered on the unperturbed sites (random area sampling), so the models interpolate between ordered for $\varepsilon=0$ and disordered for $\varepsilon=1$. All models have porosity $\varphi=1-\pi/8\sqrt{3}\approx 0.773$ and mean pore-throat size $\bar{\lambda}=a$, independent of $\varepsilon$. Fluctuations in the pore-throat sizes obey $\sigma_{\lambda} \approx 0.182\bar{\lambda} \varepsilon $ according to the approximate analytical throat-size probability density function derived in appendix~\ref{sec:app_throats}. We fabricated five different porous cells with $\varepsilon \in \{0.00, 0.25, 0.50, 0.75, 1.00\}$, spanning a full range of heterogeneity parameters from completely ordered ($\varepsilon=0$) to completely random ($\varepsilon=1$). These cells have the same pore-space topology for all $\varepsilon$ as characterized by the Euler characteristic \citep{feder_physics_2022}.
Thus, the $\varepsilon$ parameter modifies the pore-size heterogeneity without changing the pore-space topology.
\subsection{Particle tracking velocimetry and velocity field preparation}
Velocity fields were obtained in these models using particle tracking velocimetry (PTV), which reconstructs velocity fields from image sequences of tracer particles in the flow \citep{morales_stochastic_2017, souzy_velocity_2020}.
Unlike particle image velocimetry (PIV), which correlates image windows to produce gridded fields \citep{roman_particle_2016, lindken_microparticle_2009}, PTV achieves sub-pixel resolution \citep{kahler_resolution_2012}, making it well-suited to probe near-boundary velocities in porous media.
Green fluorescent particles of emission wavelength $\approx \SI{520}{\nano\meter}$, diameter $\approx$ \SI{20}{\micro\meter}, and density $\SI{1.00}{\gram\per\centi\meter\cubed}$ were dispersed in deionized water containing \SI{0.01}{\percent} by mass of Tween 80 surfactant using an ultrasonic bath. The suspension was pumped through the porous model at flow rate $q=\SI{0.5}{\milli\liter\per\minute}$ with a syringe pump, which in the model geometries gave an average flow velocity $\bar{u}_{\text{avg}} = \SI{0.1}{\milli\meter\per\second}$ and a Reynolds number $\Rey = 5\times10^{-4}$, low enough to neglect the inertia term in the Navier--Stokes equations. The Stokes number was $\Stokes = 6\times10^{-6},$ small enough to consider that the instantaneous particle velocities closely approximate the underlying flow velocities \citep{lindken_microparticle_2009}. The cell was illuminated from below by a blue light-emitting diode (LED) panel with a wavelength range $450 \pm \SI{30}{\nano\meter}$, with long wavelengths screened by a short-pass filter gel (Lee 126). The fluorescence of moving particles was recorded through a bandpass filter (wavelength range $520 \pm \SI{10}{\nano\meter}$) with a 5-megapixel camera (Allied U511M) operating at \SI{10}{\hertz}, with 8-bit intensity resolution and $\SI{24}{\micro\meter/pixel}$.

For each porous model, a ten-minute-long image sequence $I_t$ was obtained during fluid flow. A background image was formed by taking the minimum of each pixel across the sequence, $B=\min(I_t)$, and a foreground sequence showing only the moving particles was formed as $F_t = I_t-B$. A solid mask $M$ was obtained by thresholding the background image to isolate the cylinders. PTV was conducted on the background-subtracted images ($F_t$) using the open-source software TracTrac \citep{heyman_tractrac_2019}. TracTrac identifies particle centers with subpixel precision by thresholding the result of a difference-of-Gaussian filter, and it associates identified particles across temporal frames into trajectories using a Kalman-filtering approach. The analysis of each ten-minute image sequence provided around $10^6$ trajectories, and these were further filtered by their tortuosity and length, giving around $2\times10^5$ cleaned particle trajectories per experiment.

The particle trajectories were binned into velocity fields $\v{u}=(u_x, u_y)$ on the same \SI{24}{\micro\meter} grid as the original experimental images.
For each pixel, all trajectories that crossed it at some time were identified, and the crossing velocities of these trajectories over the pixel were recorded.
Note that our measurements sample tracers within a three-dimensional flow cell, whereas our target is the mid-plane velocity.
The upper 75th percentile of crossing velocity observations was averaged, ensuring that the recorded velocity at each pixel approximates the channel mid-plane value, where the velocity is maximal.
The 75th percentile was chosen as a compromise between data quantity and precision, and we found the overall structure of the velocity field was insensitive to the choice.
The velocity components were set to zero on the solid mask $M$.
Pixels which had no velocity observations were interpolated from neighboring values using bi-harmonic inpainting from scikit-image \citep{walt_scikitimage_2014}.

The PTV-derived velocity fields were divergence corrected using a custom-built algorithm adapted from \citet{wang_weighted_2017} for application to porous media flows. Divergence correction was found to be essential for the calculation of fluid stretching. From each field $\v{u}$, we find the divergence-corrected field $\v{u}'$ that minimizes $\int d^2\v{x} ||\v{u}'-\v{u}||^2$ subject to the constraint that $\nabla \cdot \v{u}'=0$ on all fluid cells.

\citet{wang_weighted_2017} provided an efficient solution of this constrained optimization problem, but their approach acts on both fluid and non-fluid cells, so it does not account for the no-slip condition on solid boundaries. Therefore, we implemented the constrained problem using a brute-force approach on a GPU using the Cooper package, built on PyTorch \citep{gallego-posada_cooper_2025, paszke_pytorch_2019}.
Our algorithm exactly enforces zero velocities on the solid mask $M$. This means that the velocity field naturally adopts the appropriate near-boundary behavior required to satisfy both incompressibility and the no-slip condition, namely that the velocity component normal to the solid surface increases quadratically while the tangential component increases linearly with distance from the wall \citep{leal_advanced_2007}.
Our porous media divergence-correction implementation is freely available on Github \citep{pierce_deldotnot_2025}.
We applied the algorithm until the divergence change per iteration fell below $10^{-4}\bar{u}/a$.
We confirmed that the flow and elongation statistics of interest were insensitive to increases in the allowed divergence residual.
The resulting divergence-free velocity fields $\v{u}'=(u_x',u_y')$ are used in the following, dropping the primes for convenience.
To benchmark the PTV measurements, the flow was simulated using finite-volume method for the case $\varepsilon=0.50$ and compared with the divergence-corrected experimental flow field, showing qualitatively identical flow structure with errors less than $10\%$ in the measured velocity components.

\subsection{Elongation statistics of material lines}
\label{sec:shearmethods}
The stretching of infinitesimal fluid line segments in steady two-dimensional flows has been calculated in closed form by \citet{dentz_coupled_2016}. Their solution relies on transforming the general formulation of fluid stretching \citep[e.g.,][]{cocke_turbulent_1969, drummond_turbulent_1990} into a Protean coordinate system, which is an evolving orthogonal basis that moves with the local flow velocity and rotates its basis vectors parallel and perpendicular to the local flow direction at all times \citep{adachi_calculation_1983}. An infinitesimal filament (i.e., a lamella) in a two-dimensional steady flow is represented by the vector $\v{z}=(z_\parallel,z_\perp)$, with components parallel and perpendicular to the streamline, respectively. This vector evolves according to
\begin{subequations}
\begin{align}
z_\parallel(t) &= \frac{u(t)}{u(0)}\left[ z_\parallel(0)+z_\perp(0)\int_0^t \frac{\sigma(t')}{[u(t')/u(0)]^2}dt'\right]
\label{eq:stretch_par}
\\
z_\perp(t) &= \frac{u(0)}{u(t)}z_\perp(0),
\label{eq:stretch_perp}
\end{align}
\label{eq:stretch}
\end{subequations}
where $u(t)$ and $\sigma(t)$ are respectively the streamwise velocity and the streamwise shear rate (i.e., the shear acting across a streamline) along the segment's trajectory. Note that a change of variables can also express these equations in terms of the stream-line arc-length $s$ \citep{dentz_coupled_2016}.
The elongation is given by $\rho(t) = |\v{z}(t)|/|\v{z}(0)|$, which according to equation\ \eqref{eq:stretch} originates from both velocity fluctuations and accumulated shear. The Protean-frame stretching theory of equation\ \eqref{eq:stretch} has also been extended to three-dimensional steady flows \citep{lester_fluid_2018, lester_mixing_2025} and unsteady flows \citep{lesterFluidDeformationRandom2026}. We refer to $K(t)=\sigma(t)/[u(t)/u(0)]^2$ as the deformation kernel and $D(t)=\int_0^t K(t') \, dt'$ as the deformation integral.

The deformation integral in equation \eqref{eq:stretch} is the source of algebraic stretching in two-dimensional steady flows \citep{dentz_coupled_2016}.
The integral kernel reveals a nonlinear coupling between the velocity and shear in determining fluid stretching.
The deformation kernel weights the streamwise shear by two factors of inverse velocity, which can be understood as follows. The first factor of $1/u(t)$ represents the persistence of stretching, ensuring that streamwise shear produces additional deformation when sustained for longer. The second factor of $1/u(t)$ represents mass conservation: in the incompressible flow, deceleration compresses filaments in the streamwise direction and extends them transversely (equation\ \eqref{eq:stretch_perp}), so that shear acts on the extended transverse component $z_\perp(t') = u(0)z_\perp(0)/u(t')$ \citep{winter_modelling_1982}. Mass conservation similarly explains the $u(t)/u(0)$ pre-factors in equations\ \eqref{eq:stretch}.

The stretching theory of equations\ \eqref{eq:stretch} is challenging to apply in practice, since the integrals involve velocities and velocity gradients along Lagrangian trajectories, and these quantities rarely have analytical solutions. However, in porous media, we can expect the ratio $\sigma/u^2$ to peak near solid boundaries where the velocity is low and the shear is high. The ratio is comparatively negligible elsewhere, since shear decreases while velocity increases with distance from walls. Later, we will exploit this wall-localized character of $\sigma/u^2$ in porous domains to explain stretching in porous media based on the frequency with which the solute plume encounters solid boundaries.

We calculate the elongation statistics of material lines using the strip method \citep{meunier_diffusive_2010} applied to the measured flow fields.
In each flow, we initialize a line of particles oriented approximately perpendicular to the mean flow and placed as far from nearby solid boundaries as possible. Neighboring particles on the line are initially separated by bonds of length $\delta \ell(0) = 0.05a$ carrying arbitrary mass $1$.
These particles are transported in the flow by solving $d\v{x}(t)/dt=\v{u}[\v{x}(t)]$ with an RK4 scheme and cubic interpolation on the velocity fields \citep{giordano_computational_2006}. The bond lengths evolve through time to $\delta \ell(t)$---generally different for each bond, and the elongation of each bond is calculated as $\rho(t) = \delta\ell(t)/\delta\ell(0)$. Whenever neighboring particles become separated by more than $2\delta \ell(0)$, the bond is subdivided and another particle is inserted at its midpoint. Subdivided bonds inherit the elongation $\rho$ of their parent bond with half its mass, such that the mass carried by a bond is $2^{-n}$, with $n$ being the total number of subdivisions applied. Refinement conserves the overall mass and elongation of the line. The resulting elongation distribution $P(\rho,t)$ and moments $\langle \rho(t)^k \rangle$ are calculated by weighting the values of $\rho$ across all bonds by their masses.

\section{Flow velocimetry and fluid elongation results}
\subsection{Velocity statistics for different solid heterogeneity}

To evaluate fluid elongation as a function of medium heterogeneity, we first obtain the velocity and shear fields calculated from the particle-tracking velocimetry in the porous media cells.
The PTV data provide the vector field $\v{u} = (u_x,u_y)$, from which we calculate the streamwise velocity field $u = |\v{u}|$ and the flow direction field $\v{u}/u$. To obtain the streamwise shear rate, we use a minimal approach based on rotating the velocity gradient tensor into the streamline frame rather than following the full Protean-frame transformation.
We compute the velocity gradient tensor $\tensor{L} = \nabla \v{u}^\top$ and assemble an orthogonal matrix $\tensor{R}=[\vhat{t},\vhat{u}]$ that rotates this tensor into coordinates with one axis $\vhat{t}=(u_r,u_\theta)/u$ tangential to the flow and another axis $\vhat{n}=(-u_\theta,u_r)/u$ normal to it.
The velocity gradient tensor transforms into the streamline frame via $\tensor{L}' = \tensor{R}^\top\tensor{L}\tensor{R}$.
This transformation aligns the reference axes with the flow direction at each point, but it does not transport them along with the flow, so it is not a full Protean transformation.
Still, $\tensor{L}'$ conveniently isolates the streamwise shear as the sum of off-diagonal components, $\sigma = L'_{12}+L'_{21}$ \citep{dentz_coupled_2016}.
\begin{figure}[htb]
    \centering
    \includegraphics[width=\linewidth]{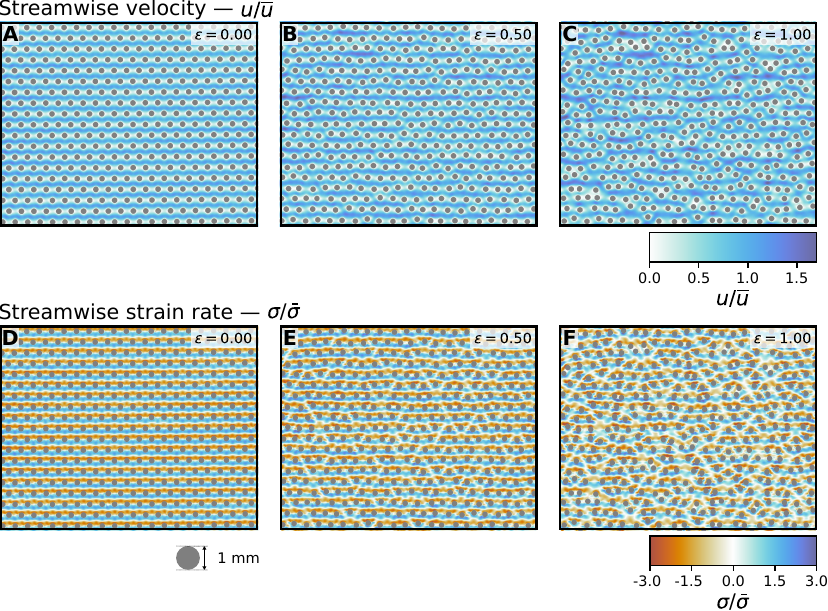}
    \caption{The streamwise velocity and shear rate fields are displayed for three of the observed heterogeneity levels. In (A), the ordered model displays two nonzero velocity modes, the channel-like pathways where the velocity is high, and the wake-like pathways behind the grains where the velocity is low. Comparison of the velocity fields in (A), (B), and (C) shows an increase in velocity heterogeneity with disorder that relates to a blending of the high and low-velocity pathways visible in (A). Likewise, the shear rate $\sigma$ becomes more spatially heterogeneous with increasing disorder parameter $\varepsilon$. Shear magnitudes are typically large on the side walls of cylinders and drop to zero in the widest pores where the velocity is high.}
    \label{fig:velocityFields}
\end{figure}

We scale the measured velocities by the imposed mean velocity $\bar{u}=q/(S\varphi)$, where $S= \SI{2.25}{\centi\meter\squared}$ is the cross-sectional area, and we scale the shear rates by the inverse of the advection time, $\bar{\sigma}=1/t_a=\bar{u}/a$. The scaled streamwise velocity and shear-rate fields are displayed for a representative subset of the measured heterogeneity levels in figure~\ref{fig:velocityFields}.
Disorder increases the velocity variance in figure~\ref{fig:velocityFields}A-C and organizes the flow structure into preferential paths that increasingly carry more of the total flux as $\varepsilon$ increases. Two different fundamental flow structures are visible in the velocity fields, especially in the ordered case $\varepsilon=0$. The first structure is the high velocity channel-like flows along the open flow pathways following symmetry directions of the underlying cylinder arrangement (typically in blue, see figure~\ref{fig:velocityFields}A). The second structure is the obstructed flow pathways sheltered between cylinders (typically in white). Increasing disorder blends these two fundamental flow structures and diversifies the distribution of flow directions.

The streamwise shear rate similarly shows two modes that blend together with increasing heterogeneity, see figure~\ref{fig:velocityFields}D-F. The shear rate changes sign in the transverse direction across the widest flow pathways. Shear-rate magnitudes are typically highest along cylinder walls, especially near the sides of cylinders where the flow velocity is largest in the near-wall regions. Disorder increases the variability and range of shear-rate values.
\begin{figure}[htb]
    \centering
    \includegraphics[width=\linewidth]{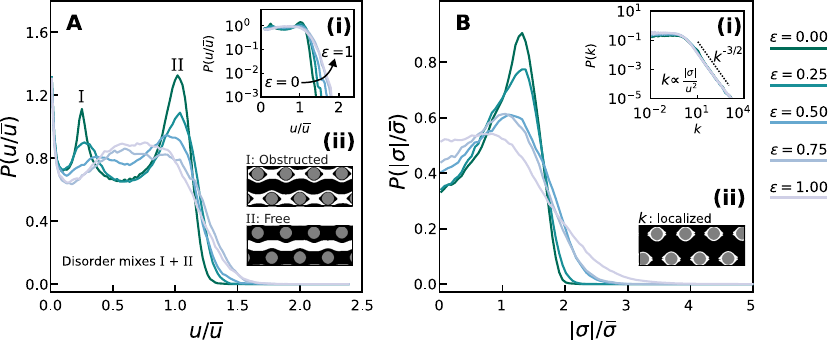}
    \caption{Panel (A) displays the velocity statistics for different model heterogeneities (denoted by colors). Ordered media display two peaks in the velocity probability density function associated with obstructed- and free-flow pathways. Inset (i) shows the exponential tail of $P(u)$ that broadens with $\varepsilon$. Inset (ii) highlights the obstructed- and free-flow modes in the $\varepsilon=0$ velocity distributions; these modes increasingly blend together as $\varepsilon$ increases. Panel (B) shows the streamwise shear-rate statistics. The free-flow mode $\sigma\approx \overline{\sigma}$, visible for $\varepsilon=0$, progressively transfers to low and high $\sigma$ as disorder increases. Inset (i) displays the probability distribution of the dimensionless deformation kernel $k = (|\sigma|\bar{u}^2)/(u^2\bar{\sigma})$ (equations\ \eqref{eq:stretch}). The distribution of $k$ displays a power-law decay over more than three decades, regardless of the disorder level. This scaling is derived for the flow around an isolated cylinder in appendix\ \ref{sec:app_kernel}. Inset (ii) shows locations where $k>50$, indicating that stretching potential is concentrated in thin margins near solid boundaries. A more detailed image of the $k$-field is available in figure~\ref{fig:concept}A.}
    \label{fig:velocityStatistics}
\end{figure}
To quantify the velocity and shear characteristics as a function of medium heterogeneity, we calculate the probability density functions of streamwise velocity and shear from the measured PTV data in figure~\ref{fig:velocityStatistics}. Velocity statistics show a multimodal structure with two non-zero modes, coincident with the obstructed- and free-flow modes described above (figure~\ref{fig:velocityStatistics}A). Heterogeneity blends these non-zero modes together as $\varepsilon\rightarrow 1$. 
The velocity distributions are approximately flat for $u\rightarrow 0$ and have an exponential tail for large $u>\bar{u}$, which becomes wider as medium heterogeneity increases, as shown in inset (i) of figure~\ref{fig:velocityStatistics}A. 
A flat velocity distribution is anticipated for $u\rightarrow 0$ due to the no-slip condition on solid boundaries \citep{deanna_prediction_2017, dentz_mechanisms_2018}.
The obstructed- and free-flow modes contributing to the distributions are visualized in inset (ii) of figure~\ref{fig:velocityStatistics}A. The upper graphic is white where $0.1\leq u/\bar{u} \leq 0.5$, corresponding to the obstructed mode, while the lower graphic is white where $0.8 \leq u/\bar{u} \leq 1.2 $, corresponding to the free mode.

The streamwise shear rate magnitude $|\sigma|$ follows a unimodal distribution as shown in figure~\ref{fig:velocityStatistics}B. The most probable value for all media is near the value $\bar{\sigma}$ defined by the mean velocity and grain size. Increasing heterogeneity broadens the shear variability and flattens the weak-shear component of the distribution ($\sigma<\bar{\sigma}$), associated with high-velocity regions away from walls. The shear distribution shows an exponential-like tail for large values, associated with low-velocity regions near walls.

While the shear rate and velocity fields have a complicated spatial structure for all heterogeneity levels (figure~\ref{fig:velocityFields}), it is the combination $\sigma/u^2$ which produces stretching in the deformation integral of equation\ \eqref{eq:stretch}a, and this quantity shows relatively simple behavior. Inset (i) of figure~\ref{fig:velocityStatistics}B plots the probability density function of $k = K/\bar{K}$, where $\bar{K}=\bar{\sigma}/\bar{u}^2$ is a characteristic deformation kernel value. We observe a similar $P(k)$ for all heterogeneity levels, with a flat distribution for small values of $k \lesssim 1$ and an identical power-law decay $P(k)\sim k^{-3/2}$ over more than three decades for the large values of $k$ associated with near-wall regions.
These large-$k$ values are expected to dominate the deformation integral in equations\ \eqref{eq:stretch}. Inset (ii) of figure~\ref{fig:velocityStatistics}B indicates the wall-localized character of these large deformation kernel values.
Thresholding the fields where $k>50$ demonstrates that the power-law component of the deformation kernel distribution is exclusively located in thin shells around solid grains, suggesting that the near-wall flow will be primarily responsible for fluid stretching in our porous media.
\subsection{Impact of heterogeneity on fluid stretching}
\label{sec:hetero}
\begin{figure}[!ht]
    \centering
    \includegraphics[width=\linewidth]{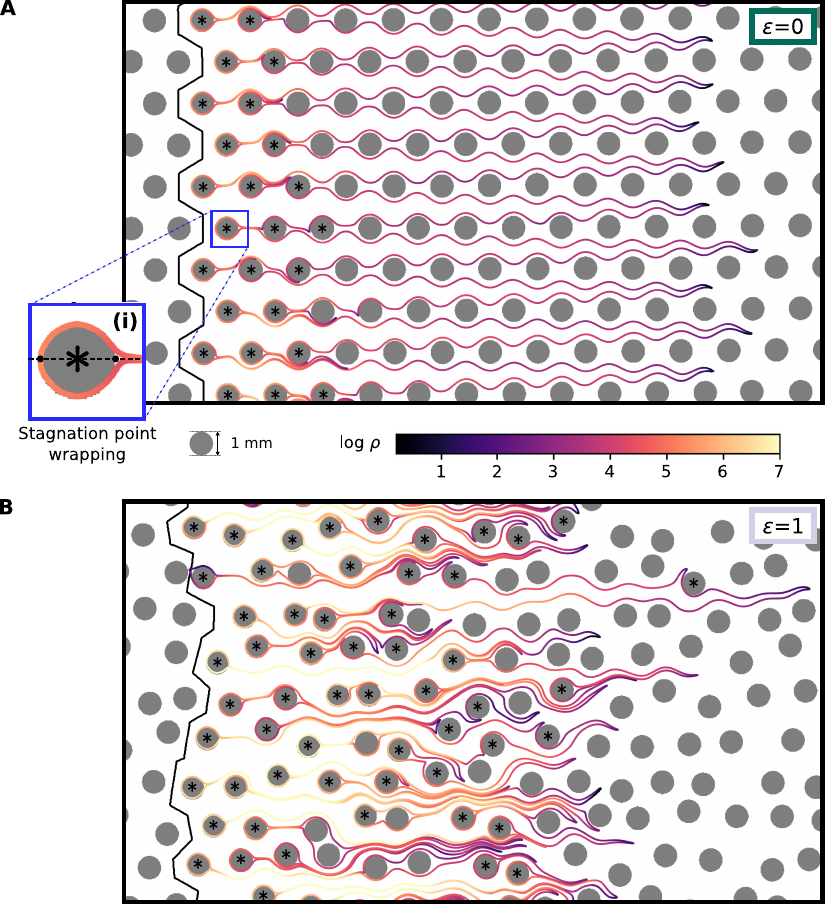}
    \caption{Material line elongation for completely ordered ($\varepsilon=0$) and disordered ($\varepsilon=1$) models. Each gray cylinder has diameter $a = \SI{1}{\milli\meter}$. The number of accumulated wraps of the plume over cylinders is a key control over the stretching statistics. Wraps are highlighted with the symbol ($\ast$). In ordered media, the number of wraps tends toward a constant (Panel A), whereas in disordered media, the number of wraps steadily accumulates with time (Panel B). In all cases, an overall spatial organization of the stretching is visible, with the earliest stagnation points to be wrapped originating the most highly-stretched sections of the plume.}
    \label{fig:plumes}
\end{figure}

The elongation of an initial material line from the advective strip calculations is displayed in figure~\ref{fig:plumes} for two representative solid geometries. The initial material line is displayed as a black line, while the stretched line at $t \approx 3 \tau_a$ is displayed with a color scale that represents the local elongation.
We confirmed that stretching statistics were insensitive to the layout of the initial line by computing $\langle\rho\rangle(t)$ for five different initial lines at each $\varepsilon$.
Averaging the standard deviation of $\langle\rho\rangle(t)$ over $t/t_a > 5$ provided a 1.3\% uncertainty in the mean elongation due to the initial line placement.
This insensitivity is expected since the line is generally aligned perpendicular to the local flow, while stretching is concentrated in the parallel direction (equation \eqref{eq:stretch_par}).
Ordered porous media show relatively less diversity in the elongation distributions, and there is a clear spatial organization of the elongation from most-stretched on the left (near the initial configuration) to least-stretched on the right. The plume wraps around the first cylinders it encounters, near its initial position, then weaves through the widest pore throats and mainly avoids subsequent cylinder encounters. When additional wrapping events do occur, they appear to arise from asymmetries in the flow field from finite system size: one can observe a ``fanning-out'' of the flow from the inlet port (see figure \ref{fig:plumes}A).
This influence persists for short distances around $\sim 3a$, consistent with the screening in porous media of the otherwise long-ranged $1/r$ interactions in Stokes flows \citep{durlofskyAnalysisBrinkmanEquation1987}.

In contrast, disordered porous media show more diversity in the elongation of material lines, and wrapping of cylinders is not limited to the first cylinders the plume encounters. The plume continually encounters new cylinders as it progresses downstream, and the most elongated areas exist just downstream of the earliest cylinders to be wrapped during transport. This organization of stretching around cylinder encounters likely relates to the localization of $\sigma/u^2$ near solid boundaries. Material lines continuously accumulate contacts with wall regions in disordered porous media, whereas in ordered porous media, material lines only encounter wall regions early in transport and do not gain further contacts thereafter. This difference in the frequency of wall encounters coincides with the higher magnitude and diversity of elongation seen in the disordered media (figure~\ref{fig:plumes}).

\begin{figure}[htb]
    \centering
    \includegraphics[width=\linewidth]{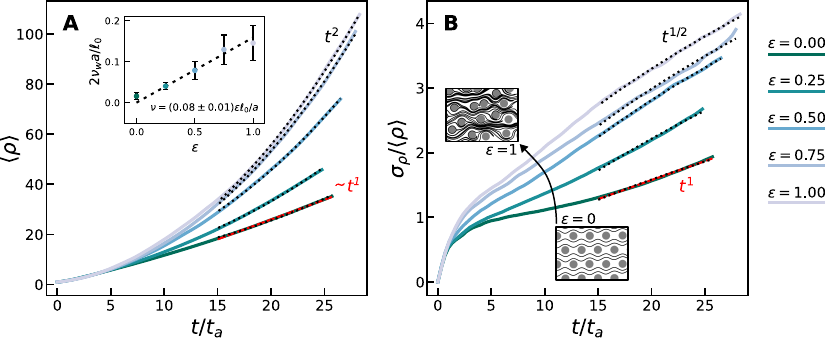}
    \caption{First two moments of the elongation for all disorder levels. In panel (A), the asymptotic mean elongation grows quadratically for disordered media, approaching linear for ordered media. 
    The quadratic growth implies that the elongation process in disordered media cannot be approximated as a simple shear flow. In panel (B), the relative variation $\sigma_\rho/\langle \rho\rangle$ of the elongation is displayed. The scaling $\sigma_\rho/\langle \rho\rangle\sim t^{1/2}$ is superimposed for disordered media, while $\sigma_\rho/\langle \rho\rangle\sim t^1$ is superimposed for ordered media. The stretching model developed in section \ref{sec:aggregates} predicts these scaling relations. The inset of panel (B) shows an increased density of neighboring filaments in disordered media, reflecting an increased propensity for lamella aggregation during mixing.}
    \label{fig:stretchingmoments}
\end{figure}

\begin{figure}[htb]
    \centering
    \includegraphics[width=\linewidth]{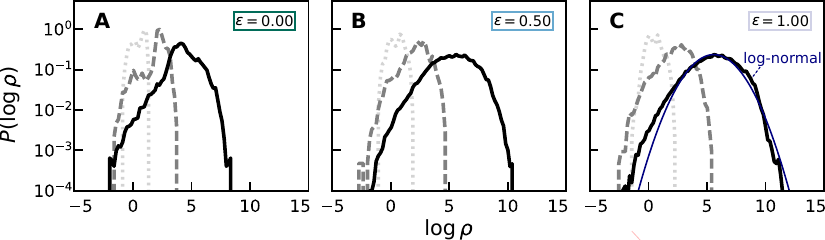}
    \caption{Probability density functions of $\log\rho$ are displayed for times $t/t_a \in \{1,5,25\}$ and different heterogeneity levels: $\varepsilon=0.00$ (A), $\varepsilon=0.50$ (B), and $\varepsilon=1.00$ (C). 
    In all cases, the observed distributions deviate from a log-normal shape. Panel (C) displays a log-normal fit for comparison.
   }
    \label{fig:stretchingpdf}
\end{figure}

To quantify the elongation statistics, we calculate the first two moments of the elongation for all fabricated porous structures. Figure~\ref{fig:stretchingmoments} displays the mean elongation $\langle \rho \rangle$ and the relative variation of the elongation, $\sigma_\rho/\langle\rho\rangle$ for the studied media. The notation $\sigma_\rho^2 = \langle \rho^2\rangle - \langle \rho\rangle ^2$ denotes the elongation variance. In ordered media, the mean elongation scales approximately linearly with time, whereas in disordered media, the mean elongation scales quadratically with time, with a pre-factor that increases with the disorder parameter $\varepsilon$.
The distribution of $\log\rho$ is displayed in figure~\ref{fig:stretchingpdf} for three representative cases. With increasing time, the distribution broadens and its mode shifts to larger $\log\rho$. The elongation distribution is more skewed for the ordered medium, whereas for the disordered medium, the distribution more resembles a log-normal distribution, although with additional frequency at small or compressive elongations when compared to log-normal. 
\section{Random walk model for elongation in cylinder arrays}
\label{sec:stretchingModel}
This section develops a quantitative description for the observed elongation statistics in figure~\ref{fig:stretchingmoments}.
In particular, the description aims to predict the asymptotic scaling of the elongation moments $\langle \rho^k \rangle$ with time for the full range of cylinder array heterogeneities $0\leq \varepsilon\leq 1$.
In addition to depending on the degree of heterogeneity $\varepsilon$, stretching characteristics in fully ordered arrays ($\varepsilon=0$) are expected to depend on the flow orientation relative to the lattice structure; however, we only study flows aligned with the lattice here.
To develop the description of elongation, we exploit the localization of the deformation kernel $\sigma/u^2$ in the stretching equations~\ref{eq:stretch} to the near-wall regions of individual cylinders as shown in the figure~\ref{fig:velocityFields}B inset.

\begin{figure}[htb]
    \centering
    \includegraphics[width=\linewidth]{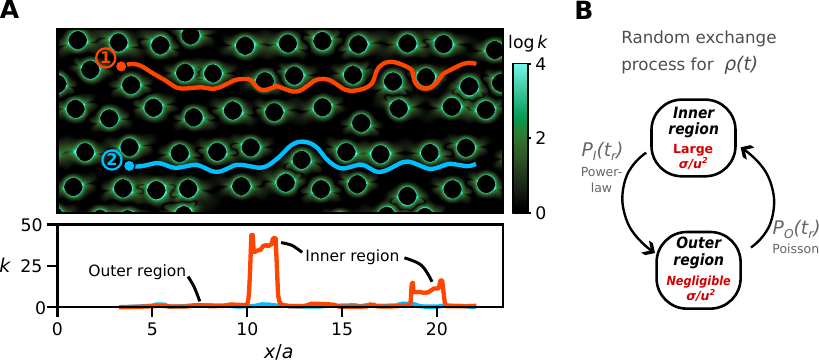}
    \caption{Proposed conceptual model for the transport of an infinitesimal lamella as random alternation between inner (near-wall) and outer (bulk flow) region flows.
    (A1) Scaled deformation kernel $k = (|\sigma|/u^2)/(\overline{|\sigma|}/\bar{u}^2)$ for $\varepsilon=1$, showing close localization of this quantity to cylinder walls, implying the deformation integral in equation~\ref{eq:stretch} grows mainly due to boundary encounters. Trajectories of two different lamellae in the flow are indicated. 
    (A2) Scaled deformation kernel sampled by the indicated trajectories. Only close encounters with cylinders result in appreciable shear. Lamella 1 is strongly stretched by two inner-region transits, while lamella 2 accumulates negligible stretching by remaining in the outer region. (B) Alternation scheme of a lamella between inner- and outer-flow states, with the residence time distributions in each state given by equations \ref{eq:outerResidenceTime} and \ref{eq:innerLayerResidenceTime}, respectively.
    }
    \label{fig:concept}
\end{figure}

We decompose the initial line into infinitesimal lamellae and represent the transport of each lamella as a random exchange process between two states: (1) ``inner regions'' of cylinders where $\sigma/u^2$ is large, corresponding to locations where the no-slip and incompressibility conditions constrain the flow-field \citep[e.g.,][]{hinch_perturbation_1991}; and (2) the ``outer region'' or bulk flow where $\sigma/u^2$ can be considered negligible.
Incorporating the transit time distributions through each of these two states will allow us to construct a multi-state random-walk description \citep[e.g.,][]{weiss_aspects_1994} for the elongation statistics of lamellae.
Figure~\ref{fig:concept} summarizes the conceptual basis of the model as random alternation between inner and outer regions; with transit time distributions $\psi_I(t_r)$ and $\psi_O(t_r)$, respectively, with stretching occurring only in the inner-region transits where the deformation kernel is large.

We analytically study the elongation produced when a material line wraps over an isolated cylinder, and we assume this sufficiently represents the elongation produced by a cylinder within a larger array to parameterize the random-walk model.
The single-cylinder analysis implies truncated power-law distributions for both the transit times and elongation increments through inner regions.
We assess the rates at which the material line wraps over cylinders from the stretching simulations.
This provides access to a medium-dependent wrap rate and suggests Poisson-distributed transit times through the outer region. 
When incorporated in a random exchange model as shown in figure~\ref{fig:concept}B, these distributions will be seen to produce anomalous scalings through time of the elongation moments which depend on medium disorder and nicely match our simulation results.

\subsection{Quantification of cylinder wrapping events}
\label{sec:wrapping}
\begin{figure}[htb]
    \centering
    \includegraphics[width=\linewidth]{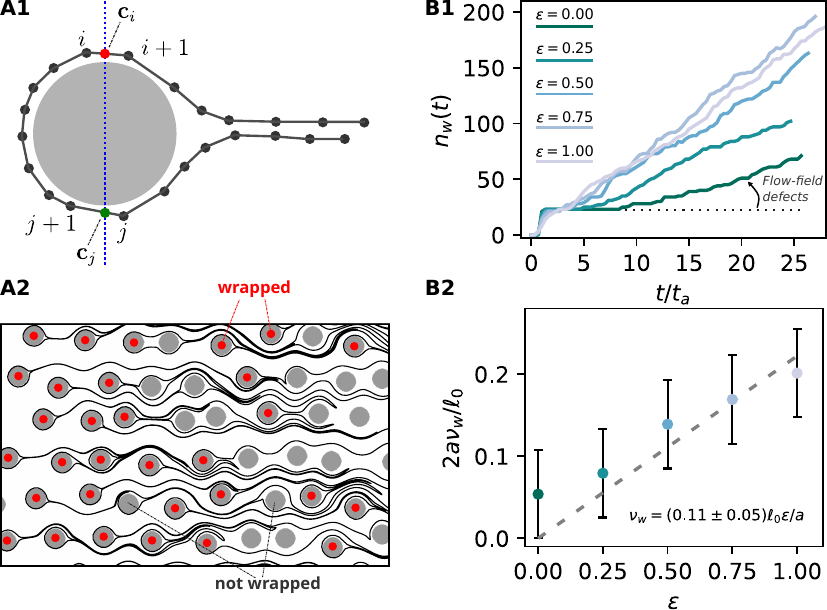}
    \caption{Evaluation of cylinder-wrapping events from stretching simulations. A1 illustrates of the operational definition of wrapping. A2 displays an elongated line with wrapped cylinders highlighted by red dots. B1 shows the cumulative number of wraps versus time, suggesting a relationship $n_w\sim \nu_wt/t_a$, generally increasing with medium disorder. Asymmetries in the flow field produce a finite-time departure of the $\varepsilon=0$ case from $\nu_w \sim \text{const.}$. B2 shows the dimensionless wrap rate $\nu_w$ scaled by the quantity $2a/\ell_0$ with initial filament length $\ell_0$. Linear fits are developed to estimate $\nu_w$ over $2\leq t/t_a \leq 10$, before finite-time effects become relevant. The departure of the $\varepsilon=0$ case from $\nu_w\sim 0$ is used as an upper error bound for all cases. In B1, the data support a linear relation $\nu_w = (0.11\pm 0.05)\varepsilon a/\ell_0$, although with large uncertainty from the definition used to detect wraps, showing an increase in wrap rate with medium heterogeneity. An improved relation is obtained with other methods in section~\ref{sec:stretchingModel}.}
    \label{fig:wrapping}
\end{figure}

Wrapping events would ideally be characterized as the bending of material lines over the no-slip stagnation points at the upstream sides of cylinders, but the detection of these events is challenging with discrete velocity fields and material lines represented as discrete points.
Instead, to identify locations where the material line wraps over cylinders, we introduce an operational definition shown in figure~\ref{fig:wrapping}A1.
The center points of all cylinders are identified with a Hough transform applied to the solid mask $M$, and rays of length $2a$ are drawn transverse to the mean flow across each center point.
The material line is represented by a chain of points $\v{x}_i$, $i=1\dots N(t)$, and a wrap is considered to occur over a cylinder at time $t$ if three conditions are met: (1) two bonds $\v{x}_i$-to-$\v{x}_{i+1}$ and $\v{x}_j$-to-$\v{x}_{j+1}$ on the line both cross the same ray; (2) the cumulative distance $s_{ij} = \sum_{k=i}^j |\v{x}_k|$ between these segments satisfies $s_{ij} < 3\pi a/4$ (less than half of the cylinder circumference plus a 50\% tolerance); and (3) the two (approximate) crossing points $\v{c}_i = (\v{x}_i + \v{x}_{i+1})/2$ and $\v{c}_j = (\v{x}_j + \v{x}_{j+1})/2$ are separated by more than a distance $a$.
This definition successfully detects cylinder wraps while avoiding false detections, as shown in figure~\ref{fig:wrapping}A2. Here, wrapped cylinders are highlighted with red dots for one timestep of the $\varepsilon=1$ simulation.

Applying the definition to quantify the number $n_w(t)$ of wrapping events versus time summarizes how the medium heterogeneity affects the frequency of encounters by the material line in figure ~\ref{fig:wrapping}A3.
To summarize how the wrap frequency $\nu_w/t_a$ depends on medium disorder $\varepsilon$, we establish a linear fit between the wrap frequency as $\nu_w = ( 0.11\pm 0.05)\varepsilon\ell_0/a$, where $\ell_0$ is the initial length of the material line (figure~\ref{fig:wrapping}B).

The case $\varepsilon=0$ reveals a secondary late-time wrapping mechanism in the rear of the plume (see figure \ref{fig:plumes}A).
The asymptotic wrap rate $\nu_w\approx 0$ is established after $t\approx 2 t_a$, but this persists only up to $t\approx 10 t_a$ when secondary wraps develop.
These additional wraps imply that the idealized ordered-lattice result only persists for a finite time, likely as a result of the fanning-out of the flow from the inlet port which breaks translation invariance in the ordered structure.
We use the departure from a constant wrap number for $\varepsilon=0$ as an upper-bound uncertainty measure for all wrap-rate observations in figure \ref{fig:wrapping}B. 

\subsection{Stretching by an isolated cylinder}
\label{sec:isolated}
\begin{figure}[htb]
    \centering
    \includegraphics[width=\linewidth]{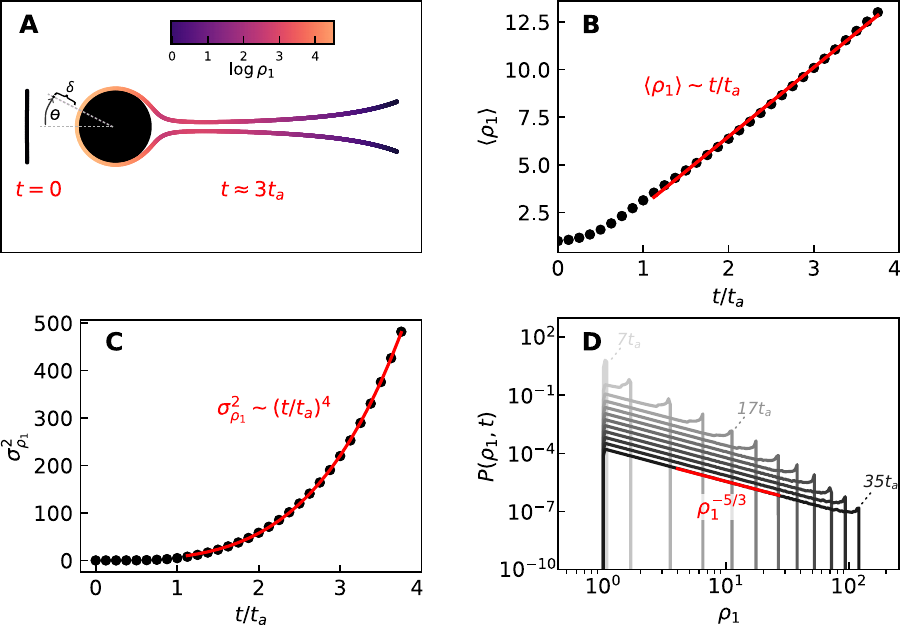}
    \caption{Numerical simulation of a material line stretching around an isolated cylinder. In (A) the line (black at $t=0$ and colored by the elongation at $t\approx 3t_a$) is transported to the right and wraps around the cylinder. The line is parameterized by a coordinate $\eta\in[-1,1]$ and an elongation $\rho_1(\eta,t)$. In (B), the mean elongation grows linearly, consistent with a concept of two branches moving past the cylinder, each with leading edge moving at velocity $\bar{u}$. In (C), the elongation variance variance grows as $t^4$. In (D), the elongation distribution $P(\rho_1,t)$ is shown evolving across evenly-spaced times (represented by the shift from gray to black), showing excellent correspondence to a truncated power-law scaling $P(\rho_1,t)\sim \rho_1^{-5/3}$ with a time-dependent maximal $\rho_1$. These distributions are vertically shifted for clarity. The upward hooks on these distributions at large $\rho_1$ originate from velocity variation near the downstream stagnation point. In all panels, red overlays show the predicted scalings from the analytical calculations in section~\ref{sec:isolated}.}
    \label{fig:singleCylinder}
\end{figure}
The observation that large deformation kernels are universally localized near solid boundaries (figure\ \ref{fig:velocityStatistics}B) suggests that encounters of lamellae with individual cylinders provide the key mechanism of stretching.
Therefore, to describe stretching in cylinder arrays, we initially study the stretching produced by flow around an isolated cylinder.

To relate to our millifluidic cells, we start from the velocity field for creeping flow around a cylinder within a Hele-Shaw cell, and we expand it for small distances from the cylinder wall.
For analytical tractability, we use the exact solution to the Brinkman equations for a cylinder in a thin gap between parallel plates \citep{pop_flow_1992,jasinski_effective_2018}.
This solution is known to provide an excellent approximation to the Stokes equations in cylinder-array geometries, where the Brinkman approximation gives maximal errors less than 2\% when compared with Stokes-flow simulations \citep{jasinski_effective_2018}.

Figure~\ref{fig:singleCylinder} displays a numerical simulation of material line stretching in the Brinkman flow around a cylinder embedded in a Hele-Shaw cell \citep{pop_flow_1992,jasinski_effective_2018}. These results were obtained by simulating the transport of $5\times10^4$ particles in the analytical Brinkman flow using the RK4 method \citep{giordano_computational_2006}. As the filament wraps the cylinder, the mean elongation scales asymptotically as $t$, while the elongation variance scales like $t^4$.
\subsubsection{Brinkman approximation and trajectories in the inner region}
To describe the single-cylinder stretching analytically, we note the Brinkman flow has both inner and outer regions: there is a Stokes-like inner region where shear is significant, and there is a Darcy-like outer region where shear can be neglected \citep{pop_flow_1992}. We non-dimensionalize the flow with the cylinder diameter $a$, the far-field velocity $\bar{u}$, and the advective time $t_a=a/\bar{u}$. 
To isolate the inner region, we expand the Brinkman flow near the cylinder wall, obtaining the velocity field
\begin{equation}
\begin{aligned}
    u_\delta &= - A\cos(\theta)\delta^2 \\
    u_\theta &=   A \sin(\theta) \delta.
\end{aligned}
\label{eq:velfields}
\end{equation}
The quantity $\delta=r/a-1/2$ represents a small dimensionless distance from the cylinder wall, and $\theta$ is the angle from the upstream stagnation point (see figure~\ref{fig:singleCylinder}A). Here, $A =4sK_1(s)/K_0(s) \approx 8.8$, where the $K_0(s)$ and $K_1(s)$ are modified Bessel functions, and $s = \sqrt{3}$ relates to the cylinder diameter and height through the permeability formula for a Hele-Shaw cell \citep{jasinski_effective_2018}. While this value of $s$ is specific to our geometry, changing the value of $s$ only modifies the pre-factors in the relations developed below.

Equation~\ref{eq:velfields} is constrained by the simultaneous need to satisfy the no-slip and incompressibility conditions, so its functional form is insensitive to the method used to derive it \citep[cf.,][]{hinch_perturbation_1991}.
Therefore, while we obtain this field using the Brinkman solution \citep{pop_flow_1992,jasinski_effective_2018}, other methods provide the same velocity field with a modified pre-factor of less than 2\%, including basis-function solutions of the Stokes equations in a gap \citep{leeStokesFlowCircular1969} and Stokes-Oseen approximation of flow around an unconfined cylinder \citep{leal_advanced_2007}.
Because the pre-factors do not affect any conclusions derived here, we can safely ignore the small differences of equation~\ref{eq:velfields} from the Stokes-flow result.

Equation~\ref{eq:velfields} implies that a particle at initial location $(\delta_0, \theta_0)$ will move along the streamline given by  $\delta(t)\sqrt{\sin\theta(t)}=\delta_0\sqrt{\sin\theta_0}$. Since this relation holds for any $t$, the quantity $b_0=\delta_0\sqrt{\sin\theta_0}$ is conserved along each streamline. Solving the system $d\delta/dt=u_\delta$ and $(1/2+\delta)^{-1}d\theta/dt = u_\theta$ provides the trajectories
\begin{equation}
\begin{aligned}
\delta &= \frac{b_0}{\text{sl}\left(b_0At + \text{sl}^{-1}(b_0/\delta_0)\right)}, \\
    \sin\theta &= \text{sl}^2\left(b_0At + \text{sl}^{-1}(b_0/\delta_0)\right).
\end{aligned}
\label{eq:trajs}
\end{equation}
The function $\text{sl}(\cdot)$ is the lemniscate sine, which is the Jacobi-elliptic generalization of the sine. The function $\text{sl}$ is periodic with period $2\varpi \approx 5.244$, analogous to $2\pi$ for the trigonometric functions.

The trajectories of equation\ \eqref{eq:trajs} break down when the particles enter the outer region near the downstream stagnation point, when $\dot{\delta}$ un-physically vanishes. This occurs for $t_r$ such that $\theta(t_r) \approx \pi-\theta_0$ (for $\theta_0>0$; the case $\theta_0<0$ follows by symmetry), which for small values of the initial angle $\theta_0$ becomes
\begin{equation}
t_r \approx  \frac{\varpi}{A\delta_0 \sqrt{|\theta_0|}}.
\label{eq:breakthrough}
\end{equation}
This relation implies that particles which started closer to the upstream stagnation point (i.e., with small $\delta_0$ and $\theta_0$) will reside longer in the inner region.
\subsubsection{Elongation through an inner-region transit}
We describe a material line in the inner region as a collection of infinitesimal lamellae at initial locations $(\delta_0, \eta\theta_0)$ indexed by 
$\eta \in [-1,1]$. The line has initial length $\ell_0 = a \theta_0$,  where $\theta_0\ll1$, and the stretching vectors $\v{z}(t|\eta)$ are conditional on $\eta$. Each lamella has $z_\parallel(0|\eta)=0$ and $z_\perp(0|\eta)=\delta\ell_0$. Here and in the following equations, the units have been restored. Lamellae will accumulate elongation as they transit through the inner region of the cylinder, since $\sigma/u^2$ is appreciable in the deformation integral of equation\ \eqref{eq:stretch_par}. Eventually, the majority of lamellae will pass into the outer region, where $\sigma/u^2$ is negligible and the deformation integral in equation \eqref{eq:stretch_par} no longer accumulates. Because most lamellae will eventually reside in the outer region of the cylinder, most of the mass eventually resides there, and the outer-region lamellae are expected to dominate the statistics. For this reason, we neglect lamellae still residing within the inner region to calculate elongation statistics in the following. This approach will be further justified later, since the derived results correspond nicely with the numerical simulations of stretching.

The flow in equation \eqref{eq:velfields} produces the constant deformation kernel in the inner region, valid to lowest order in $\delta$:
\begin{equation}
    K(t|\eta)=\frac{u_0}{\ell_0\eta}.
    \label{eq:kernel1}
\end{equation}
Appendix \ref{sec:app_kernel} describes the derivation based on the velocity gradient tensor in flow-aligned coordinates. Using equation~\eqref{eq:breakthrough}, the lamella indexed by $\eta$ resides in the inner region for a timescale
\begin{equation} t_r(\eta) \approx \tau|\eta|^{-1/2},\label{eq:breakthroughtime}\end{equation}
where $\tau = t_a\varpi/\left(A\delta_0\sqrt{\theta_0}\right)$ is the time when the first lamellae break out of the inner region (those with $\eta=\pm 1$). The distribution of times spent in the inner region follows by integrating over the uniform distribution of $\eta$:
\begin{equation}
 \psi _\text{I}(t_r) = \frac{2\tau^2}{t_r^3}\mathcal{I}(\tau\leq t_r),
    \label{eq:innerLayerResidenceTime}
\end{equation}
where $\mathcal{I}(\cdot)$ is an indicator function.
Lamellae in the outer region at time $t$ are those with $t_r(\eta)<t$, or alternatively $\eta_\text{min}<|\eta|\leq 1$, where
\begin{equation}
\eta_\text{min} = (\tau/t)^2.
\end{equation}
The deformation integral carried by outer-region lamellae is formed by multiplying the constant deformation kernel of equation\ \eqref{eq:kernel1} by the residence time over which shear was accumulated in the inner region:
\begin{equation}
    D(\eta)=\frac{u_0 \tau}{\ell_0|\eta|^{3/2}}\text{sgn}(\eta). \label{eq:deformationintegral}
\end{equation}
By assumption, all lamellae were initially perpendicular to the flow, so equations\ \eqref{eq:stretch} provide the elongation $\rho_1(t|\eta) =  [u(t) /u_0]|D(t_r)|$, where the subscript ``$1$'' denotes stretching produced by an individual cylinder. Once a lamella exits the inner region, its deformation integral $D$ is assumed frozen at the value accumulated during the inner-region transit, while its velocity $u(t)$ fluctuates independently in the outer region. Therefore, the averages split as $\langle \rho_1^q \rangle =  \langle( u/u_0)^q\rangle \langle |D|^q\rangle$. Assuming outer-region lamellae sample the mean flow velocity of the porous medium ($\bar{u}$) with narrow and symmetric fluctuations, one obtains $\langle (u/u_0)^q \rangle \approx (\bar{u}/u_0)^q$.
Accordingly, the elongation carried by the lamellae in the outer region is
\begin{equation}
    \rho_1(t_r) = \rho_\ast(t_r/\tau)^3.
    \label{eq:stretch1}
\end{equation}
Here, $\rho_\ast=\bar{u}\tau/\ell_0$ is the elongation of the first lamellae to exit the inner region ($\eta=\pm1$).
Note that equation~\ref{eq:breakthroughtime} relates $\eta$ and $t_r$, so we can equivalently write $\rho_1$ as a function of the lamella index $\eta$.
Equation \eqref{eq:stretch1} provides all elongation moments by integrating over the outer-region lamellae:
\begin{equation}
\begin{aligned}
 \langle \rho_1(t)^q\rangle &= \int_{\tau}^{t} \left[\rho_\ast(t_r/\tau)^3\right]^q p(t_r) \, dt_r \simeq \frac{2\rho_\ast^q}{3q-2} \left(\frac{t}{\tau}\right)^{3q-2}.
\end{aligned}
\end{equation}
This formula provides the mean elongation for $q=1$:
\begin{equation}
\langle\rho_1(t)\rangle \simeq  \frac{2\bar{u}t}{\ell_0}.
\label{eq:stretchsinglecyl}
\end{equation}
This result describes two branches extending downstream from the cylinder, each moving with velocity $\bar{u}$, which matches the numerical simulations in figure~\ref{fig:singleCylinder}B. Likewise, $q=2$ provides the elongation variance ($\sigma_{\rho_1}^2=\langle\rho_1^2\rangle-\langle\rho_1\rangle^2$):
\begin{equation}
    \sigma_{\rho_1}^2  \simeq \frac{1}{2}\left(\frac{\bar{u}\tau}{\ell_0}\right)^2\left(\frac{t}{\tau}\right)^4,
    \label{eq:varsinglecyl}
\end{equation}
consistent with the numerical variance scaling in figure~\ref{fig:singleCylinder}C. Under the same assumptions, we can compute the distribution of the elongation from a single cylinder at times large enough that most lamellae are in the outer region. The integral $P(\rho_1',t) = \int_{-1}^1\frac{d\eta }{2}\delta(\rho_1'-\rho_1(t_r(\eta)))$ provides
\begin{equation}
    P(\rho_1,t) \simeq \frac{2}{3}\left(\frac{\bar{u}\tau}{ \ell_0}\right)^{2/3}\rho_1^{-5/3},
    \label{eq:powerlaw}
\end{equation}
over the domain $\rho_\ast\leq \rho_1 \leq \rho_\ast(t/\tau)^3$.
This result is verified in figure\ \ref{fig:singleCylinder}D, where it is shown to correspond nicely to the numerical simulations.
We also confirmed that the upper truncation values in the distributions of figure\ \ref{fig:singleCylinder}D increase as $t^3$ in accord with the domain of equation\ \eqref{eq:powerlaw}.
The distribution of equation\ \eqref{eq:powerlaw} demonstrates that both the elongation and transit-time increments over an isolated cylinder follow truncated power-law distributions.
The truncation arises because the maximum elongation at a given time is bounded by the maximum transit time, which is given by the observation time.
Because of this truncation, all moments of equation\ \eqref{eq:powerlaw} converge at any finite time, despite the heavy-tailed decay of the elongation distribution.

\subsection{Stochastic representation of fluid stretching in porous media} \label{sec:aggregates}
We now formulate a stochastic model for fluid deformation that represents the successive deformation events produced by the encounters of the material line with cylinders.
The key model concepts are illustrated in figure~\ref{fig:concept}.
A material line is divided into infinitesimal lamellae, and each lamella undergoes a random alternation between inner and outer regions.
Transit time distributions $\psi_I(t_r)$ and $\psi_O(t_r)$ characterize the times lamellae spend in inner and outer regions, respectively.  
Elongation is considered to accumulate in inner regions only, since the deformation kernel is appreciable there and negligible elsewhere (see figure~\ref{fig:concept}A),.
The amount of elongation developed by an inner-region transit is considered as a non-linear function of the transit time, further specified below.

During flow through cylinder arrays, material lines wrap around cylinders as they stretch, and these wraps accumulate differently in ordered and disordered media (see figure~\ref{fig:plumes}).
Earlier we noted that the deformation kernel distribution has an identical ($K^{-3/2}$) power-law tail for all heterogeneity levels (see inset (i) of figure\ \ref{fig:velocityStatistics}B).
In appendix\ \ref{sec:app_kernel}, flow around an isolated cylinder is shown to produce the same power-law scaling.
Motivated by this observation, we assume any cylinder in the porous medium produces the same stretching as the isolated cylinder in section~\ref{sec:isolated}).
This representation assumes that cylinders in an array all retain single-cylinder characteristics, so it may overlook some diversity in the stretching process.
In particular, we should expect the single-cylinder approximation to break down when separations between cylinders become small compared with the typical size of regions where $\sigma/u^2$ is appreciable.
These regions are shown in figure~\ref{fig:concept}A1.
We will show that this single-cylinder assumption successfully describes the asymptotic elongation statistics in the dilute cylinder arrays considered here, but the quality of the approximation may be worse in dense arrays.

In disordered media, material lines progressively wrap around cylinders during transport at a constant rate $\nu_w/t$ which depends on medium heterogeneity (see figure~\ref{fig:wrapping}B).
Since the porous structures considered here are without long-ranged correlations and wraps accumulate at a constant rate, a Poisson model is suggested for the outer-region transit time distribution: 
\begin{equation}
    \psi_O(t_r)=\frac{\nu_w}{t_a}e^{-\nu_w t/t_a}.
    \label{eq:outerResidenceTime}
\end{equation}
This equation contains an implicit dependence on medium heterogeneity, since $\nu_w$ grows with $\varepsilon$ as shown in figure~\ref{fig:wrapping}B.

\subsection{Elongation in ordered media with flow along a symmetry axis}
\label{sec:ordered}

In ordered media with the flow oriented along a symmetry direction, material lines wrap around the first cylinders they encounter, but subsequent wraps occur only as a finite-time effect due to flow-field imperfections as shown in figure~\ref{fig:plumes}A.
The initial line wraps over a number $\ell_0/a$ of cylinders with line segments of size $a$, so the overall elongation distribution is that of an isolated cylinder (section~\ref{sec:isolated}) with $\ell_0 \rightarrow a$, giving the mean $\langle \rho \rangle \simeq {2t}/{t_a}$ and variance $\sigma_{\rho}^2 \simeq \tfrac{1}{2}\left({\tau}/{t_a}\right)^2\left({t}/{\tau}\right)^4 \sim (t/t_a)^4.$
Accordingly, flow along a symmetry axis of ordered media produces mean elongation like a shear flow with an effective shear rate $2\bar{\sigma}$.
The factor of $2$ comes from the two branches formed when the material line wraps each cylinder.
The case of flow not oriented along a symmetry axis of an ordered medium may produce quadratic elongation, similar to disordered media, although this case is not studied here.

\subsection{Random-walk description of elongation in disordered media }

To set up the random exchange process for elongation illustrated in figure~\ref{fig:concept}, we use a multi-state random walk approach \citet{weiss_aspects_1994}.
In our case, the multiple ``states'' are the inner and outer regions which have distinct transit-time and deformation statistics (see figure~\ref{fig:concept}).
Multi-state walks can be formulated using three main elements \citep{weiss_aspects_1994, pierce_back_2020, zaburdaev_levy_2015}:
\begin{enumerate}
\item  \textit{Propagators through states} (denoted $r_i$ or $R_i$) are probability densities for the time and elongation increments produced by a transit through a state $i\in\{O,I\}$. 
\item \textit{Fluxes into states} (denoted $j_i$) represent transition probability densities between one state and another at each elongation and time.
\item \textit{Probability densities in states} (denoted $p_i$) measure the likelihood of a lamella being found in state $i$ with a given elongation and time.
\end{enumerate}
These quantities can be linked through renewal equations, which serve as a generalized Master equation, to describe the elongation dynamics as a stochastic process. 
This process turns out to be non-Markovian due to the truncated power-law transit-time distribution in inner-regions.

We now define the three elements of the model, starting with the propagators.
The propagators have both ``complete'' and ``incomplete'' forms.
The complete propagators (denoted $R_i$) describe the probability density that a lamella elongates by $\rho$ in a time $t$ when a transit ends.
These involve the transit-time densities $\psi_i(t)$ given in equations~\eqref{eq:innerLayerResidenceTime} and \eqref{eq:outerResidenceTime}.
Conversely, the incomplete propagators ($r_i$) describe the probability density that a lamella elongates by $\rho$ in a time $t$ in a transit which is still ongoing.
These involves the transit-time survival probabilities $\int_t^\infty \psi_i(t')dt'$.
For complete transits, we set
\begin{equation}
\begin{aligned}
    r_O(\rho,t) &= \delta(\rho) \exp(-\nu_wt/t_a)\nu_w/t_a,\\
    r_I(\rho,t)&=\mathcal{I}(t\geq \tau)\delta(\rho-\rho_\ast(t/\tau)^3)2\tau^2/t^3,
\end{aligned}
    \label{eq:complete}
\end{equation}
using the assumed exponential wrap-time distribution of equation \eqref{eq:outerResidenceTime} for $O$ and  the single-cylinder equations~\eqref{eq:innerLayerResidenceTime} and \eqref{eq:stretch1} for $I$.
For incomplete transits, we set 
\begin{equation}
\begin{aligned}
    R_O(\rho,t) &= \delta(\rho)e^{-\nu_w t/t_a},\\
    R_I(\rho,t)&=[(\tau/t)^2\mathcal{I}(t\geq\tau)+\mathcal{I}(t<\tau)]\delta(\rho),
\end{aligned}
    \label{eq:incomplete}
\end{equation}
using the survival probabilities at time $t$, calculated from equations \eqref{eq:innerLayerResidenceTime} and \eqref{eq:outerResidenceTime}.
The indicator functions in the second equation encodes the assumption that elongation accumulates exactly at the moment when lamellae exit the inner region.

The fluxes $j_i(\rho,t)$ represent the probability density that a transit through the state $i$ begins with elongation $\rho$ at time $t$.
Assuming that all lamellae initialize in cylinder inner regions at $\rho=1$, these densities can be formulated with the renewal equations \citep{weiss_aspects_1994, pierce_back_2020, zaburdaev_levy_2015}:
\begin{equation}
\begin{aligned}
    j_O(\rho,t) &= \int_0^\rho d\rho' \int_0^t dt'\,
                   j_I(\rho',t')\, r_I(\rho-\rho',t-t'), \\
    j_I(\rho,t) &= \delta(\rho-1)\delta(t)
                   + \int_0^\rho d\rho' \int_0^t dt'\,
                     j_O(\rho',t')\, r_O(\rho-\rho',t-t').
\end{aligned}
    \label{eq:flux}
\end{equation}
Here, the $\delta$-function terms encode the initial condition, while the integrals represent all possible transitions between states.
Specifically, the integral terms sum over all paths where a lamella evolves from $(\rho', t')$ in state $j$ as it transitions into state $i$ at $(\rho,t)$.

The probability densities of $\rho$ at time $t$ in each state can be written
\begin{equation}
    p_i(\rho,t) = \int_0^\rho d\rho' \int_0^t dt'\,
                  j_i(\rho',t')\, R_i(\rho-\rho',t-t'),
    \qquad i \in \{I,O\}.
    \label{eq:prob}
\end{equation}
These equations describe a lamella that entered state $i$ at $(\rho', t')$ and remains in that state at $(\rho,t)$.
Equations \eqref{eq:flux} and \eqref{eq:prob} constitute a generalized master equation for the multistate random walk in the phase space spanned by $i\in\{I,O\}$ and $\rho$.
The elongation distribution is agnostic to which state lamellae exist in and is therefore given by the sum $P(\rho,t) = p_I(\rho,t) + p_O(\rho,t)$ over both states, which with equation \eqref{eq:prob} can be written as
\begin{equation}
P(\rho,t) = \int_0^\rho d\rho'\int_0^t dt'\Big\{ j_I(\rho',t')r_I(\rho-\rho',t-t') + j_O(\rho',t')r_O(\rho-\rho',t-t')\Big\}.
\label{eq:prob}
\end{equation}

To obtain the elongation moments, we now introduce the double Laplace transform of a function $f(\rho,t)$ as 
\begin{equation}
    \tilde{f}(g,s)=\int_0^\infty dt' \int_0^\infty d\rho' \exp(-st'-g\rho')f(\rho',t'),
    \label{eq:xform}
\end{equation} with Laplace variables $g$ and $s$ corresponding to $\rho$ and $t$, respectively.
Transforming the flux equations \eqref{eq:flux} provides
\begin{equation}
\begin{aligned}
    \tilde{j}_O &= \tilde{j}_I\tilde{r}_I,\\
    \tilde{j}_I &= e^{-g} + \tilde{j}_O \frac{\nu_w}{\nu_w+t_a s},
\end{aligned}
\label{eq:step1}
\end{equation}
where we evaluated $\tilde{r}_O(g,s) = \nu_w/(\nu_w+t_as)$ and $
   \tilde{r}_I(g,s) = 2\tau^2 \int_\tau^\infty \frac{1}{{t'}^3} \exp\left(-st' - \frac{g \rho_\ast}{\tau^3} {t'}^3\right) dt'$.
Likewise, Laplace transforming equation \eqref{eq:prob} gives
\begin{equation}
    \tilde{P} = \tilde{j}_I\tilde{R}_I + \tilde{j}_O\frac{t_a}{\nu_w + t_a s},
    \label{eq:step2}
\end{equation}
where we evaluated $\tilde{R}_O(g,s) = t_a/(\nu_w + t_a s)$ and $\tilde{R}_I(g,s) = \frac{1 - e^{-s\tau}}{s} + \tau^2 \int_\tau^\infty \frac{e^{-st'}}{{t'}^2} \, dt' $.
Combining equations \eqref{eq:step1} and \eqref{eq:step2} thus provides the exact result
\begin{equation}
\tilde{P}(g,s) =e^{-g}\frac{(\nu_w + t_a s)\tilde{R}_I+ t_a \tilde{r}_I}{t_as+\nu_w(1- \tilde{r}_I)}\label{eq:mw}
\end{equation}
for the Laplace-transformed elongation distribution.
While in principle, Laplace inversion of this equation would provide the full elongation distribution $P(\rho,t)$, the inversion is prohibitively difficult due to the integrals in $\tilde{r}_I$ and $\tilde{R}_I$ that encode the non-Markovian character of the elongation dynamics.
Note that, while equation \eqref{eq:mw} was explicitly developed for disordered cylinder arrays, it also applies to ordered arrays along symmetry directions by setting $\nu_w=0$, meaning it contains the results of section \ref{sec:ordered} as a special case.

\subsection{Moments of elongation in heterogeneous cylinder arrays}

We calculate the elongation moments $\langle \rho^q \rangle$ of order $q=1,2,\dots$ in cylinder arrays by differentiating equation\ \eqref{eq:mw} with respect to $g$.
This yields the moments of $\rho$ via the relation $\langle \rho^q\rangle = (-\partial_g)^q \tilde{P}(g,s)|_{g=0}$, an identity derived from equation \eqref{eq:xform}.
The $t\rightarrow\infty$ moments are derived from this relation using an $s\rightarrow 0$ analysis of equation \eqref{eq:mw} in appendix~\ref{sec:app_lognorm}.
This involves evaluating the integrals for $\tilde{r}_I(g,s)$ and $\tilde{R}_I(g,s)$ as $s\rightarrow 0$, incorporating these into equation \eqref{eq:mw}, and evaluating derivatives with respect for $g$ of arbitrary order as $s\rightarrow 0$.
The resulting formulas encompass both ordered (using $\nu_w=0$) and disordered (with $\nu_w\neq 0$) cylinder-array elongation moments by selecting the leading order of $t$ in equation \eqref{eq:allmoments}.
From this process, we determine the mean elongation at large times:
\begin{equation}
        \langle \rho \rangle \simeq \begin{cases} \tfrac{2t}{t_a}, & \text{ordered medium (symmetry axis)}; \\  \frac{2\nu_w a}{\ell_0}\left(\frac{t}{t_a}\right)^2, & \text{disordered medium (all } \varepsilon\neq0\text{)}.\end{cases}
        \label{eq:collectivemeans}
\end{equation}
This result shows that stretching is linear in ordered cylinder arrays when the flow aligns with a symmetry axis (section \ref{sec:ordered}), while it is quadratic in disordered arrays regardless of the amount of medium heterogeneity $\varepsilon$.
The case of an ordered medium with flow not along a symmetry axis is expected to provide elongation like a disordered medium, although we do not analyze this here.
For quadratic stretching, the pre-factor depends on the wrap rate $\nu_w$ and therefore the medium heterogeneity through the observed relation between $\nu_w$ and $\varepsilon$ in figure~\ref{fig:wrapping}B.
These scalings are overlaid on the simulated mean elongation in figure~\ref{fig:stretchingmoments}A, where they closely correspond with the data at asymptotic times.
The inset of figure~\ref{fig:stretchingmoments}A plots the pre-factor of the quadratic mean elongation against the heterogeneity parameter $\varepsilon$, showing an increasing relation, similar to the direct wrap-counting in section \ref{sec:wrapping}.

A similar calculation gives the scaling of the elongation variance:
\begin{equation}
        \sigma_\rho^2 \sim \begin{cases} t^4, & \text{ordered medium (symmetry axis)}; \\
        t^5, & \text{disordered medium (all } \varepsilon\neq0\text{)}.\end{cases}
        \label{eq:collectivevars}
\end{equation}
Together, these results provide different scalings for the relative variation of the elongation in ordered and disordered cylinder arrays:
\begin{equation}
        \frac{\sigma_\rho}{\langle\rho\rangle}\sim \begin{cases} t, & \text{ordered medium (symmetry axis)}; \\ t^{1/2}, & \text{disordered medium (all } \varepsilon\neq0\text{)}.\end{cases}
        \label{eq:collectivereldevs}
\end{equation}
These scalings are overlaid on the data in figure~\ref{fig:stretchingmoments}B, showing a good match with the stretching simulations.
Note that the disordered-media scalings also fit the ordered-media elongation statistics in figure~\ref{fig:stretchingmoments}, likely because of the flow-field asymmetries we previously noted in figure\ \ref{fig:velocityFields}A, which eventually lead to additional cylinder wrapping events in ordered media, lending late-time nonlinearity to the elongation moments.
However, the converse is not true, and the disordered-media elongation grows unambiguously as a nonlinear function of time.
The progressive wrapping of cylinders during transport is the essential ingredient to capture the observed nonlinear elongation moments in disordered cylinder arrays.

The model developed in this section is a type of coupled continuous time random walk, similar to the deformation model developed by \citet{dentz_coupled_2016}.
Building on this paradigm, we have related both the temporal statistics (equation\ \eqref{eq:innerLayerResidenceTime}) and the coupling between the timescales and elongation increments (equation\ \eqref{eq:stretch1}) to the flow around an isolated cylinder.
Additionally, we extended the random walk to have multiple states, which allowed us to integrate the temporal frequency of cylinder encounters, $\nu_w/t_a$, into the description.
Since $\nu_w$ has an increasing relation to $\varepsilon$ (section \ref{sec:wrapping}), the model developed in this section predicts faster elongation in more heterogeneous cylinder arrays, although with the same scaling exponent $\langle \rho \rangle \propto t^2$ for all disordered arrays ($\varepsilon\neq 0)$. In the discussion, we will further study the dependence of elongation on medium heterogeneity by introducing theoretical arguments to connect $\nu_w$ and $\varepsilon$.

\section{Discussion}
We have studied fluid stretching in porous media composed of heterogeneous cylinder arrays and identified different scaling relations for the elongation moments in ordered and disordered media.
To explain the observed stretching statistics, we analytically studied the stretching of a material line wrapping over a solitary cylinder, and we developed a random walk model that builds on the solitary-cylinder stretching to quantify elongation across multiple cylinder encounters.
Our approach successfully predicts the elongation moments at asymptotic times, demonstrating that the wrapping of the material line around cylinders provides the dominant mechanism of stretching in the considered porous media.
\subsection{Relation of wrap-frequency to medium structure}
\label{sec:structure}
Our stochastic description relates the pre-factor of the quadratic mean-stretching relation in disordered media (equation\ \eqref{eq:collectivemeans}) to the dimensionless frequency $\nu_w$ of cylinder wrapping.
Wrapping will occur when a streamline traced from the upstream stagnation point of a cylinder intersects the plume. In ordered cylinder arrays, these streamlines span between cylinders and do not cross the initial filament, so wraps do not accumulate. However, in disordered cylinder arrays, heterogeneity perturbs the streamlines and increases the likelihood of wrapping.

To relate the wrapping frequency to the medium structure, we assume that streamlines shift over a characteristic distance $\sigma_\lambda$, the average deviation in throat size (see appendix~\ref{sec:app_throats}), which suggests a probability $p = \sigma_\lambda/a \approx 0.182\varepsilon$ that a stagnation-point streamline shifts enough to intersect the initial plume.
Noting that the plume sweeps out $\varrho\ell_0$ cylinders in $t_a$, with $\varrho=1/(2a)$ the linear density of cylinders, we can model the dimensionless wrap rate as:
\begin{equation}
    \nu_w = \varrho\ell_0p \approx   0.09\varepsilon\ell_0/a.
    \label{eq:wrappy}
\end{equation} 
In the inset of Figure~\ref{fig:stretchingmoments}A, we fit a linear relation between $\nu_w$ and $\varepsilon$ from the pre-factors of the quadratic fit, and we determine that $\nu_w = (0.08\pm 0.01) \varepsilon \ell_0/a$, which corresponds well to the model developed here.
Using equation\ \eqref{eq:collectivemeans}, this correspondence suggests that the simple relation 
\begin{equation}
    \langle \rho \rangle = 2\varrho\sigma_\lambda(t/t_a)^2
\end{equation}
can relate the quadratic mean elongation to the medium structure in different disordered media.
In addition to inferring the wrap rate $\nu_w$ from fitting the model to the elongation observations, we also measured the wrap rate directly from the simulated plumes and found $\nu_w = (0.11\pm 0.05)\ell_0\varepsilon/a$ in section~\ref{sec:wrapping}.
Although this result has large uncertainty due to the definition used to detect wrapping events, its consistency with equation \eqref{eq:wrappy} confers additional support to the model developed here.

\subsection{Prediction of stretching in heterogeneous media}
The presented stochastic framework shares some analogies with the Lévy-walk description developed
by \citet{dentz_coupled_2016} for stretching in heterogeneous porous media at Darcy scale.
They introduced a power-law distribution $p(t)\sim t^{-1-\beta}$ of residence times and related the elongation increment and residence time as $\rho\sim t^\alpha$ by assuming a linear relationship between the velocity and shear.
While the Lévy-walk description of \citet{dentz_coupled_2016} proposed these scaling exponents phenomenologically, our framework explicitly relates them to the fluid mechanics of stretching around the isolated cylinder, evaluated in section\ \ref{sec:isolated}.
Additionally, our approach includes alternations between high- and low-velocity states to incorporate the effects of medium heterogeneity into the description.
Using the theory by \citet{dentz_coupled_2016}, we find exponents $\beta=2$ from equation\ \eqref{eq:innerLayerResidenceTime} and $\alpha=3$ from equation\ \eqref{eq:stretchsinglecyl}, which predicts mean elongation scaling as $\langle \rho \rangle \sim t^{1+\alpha-\beta}=t^2$, exactly as observed in disordered media for all $\varepsilon\neq0$ (figure~\ref{fig:stretchingmoments}A).
In contrast, the transit-time distribution is nonstationary along the symmetry directions of ordered media, so these are outside the scope of the Lévy-walk description.

In other quasi-two-dimensional porous structures besides cylinder arrays \citep[e.g.][]{hyman_stochastic_2014}, fluid stretching is likely medium-dependent, although the constrained flow properties near walls may offer a useful simplification.
Near walls, velocity distributions are typically algebraic \citep{berkowitz_anomalous_1997, kang_anomalous_2015, deanna_prediction_2017, souzy_velocity_2020}, and the shear determines the velocity as follows:
Using wall coordinates $\v{x}=(x_\perp,x_\parallel)$, the no-slip and incompressibility conditions yield velocities $\v{u}=( -\tfrac{1}{2}({\partial \sigma }/{\partial x_\parallel}) x_\perp^2,\sigma x_\perp)$ \citep{leal_advanced_2007}.
For motions predominantly along the wall, the streamwise velocity and shear thus become proportional, $u \approx  \sigma x_\perp$, which satisfies a key assumption of the Lévy-walk description \citep{dentz_coupled_2016} and provides a deformation kernel that localizes near walls as $K=\sigma/u^2\propto x_\perp^{-1}$.
Away from walls, velocity distributions are typically thin-tailed \citep{cenedese_lagrangian_1996, sederman_magnetic_2001, huang_optical_2008, matyka_powerexponential_2016} (see also figure~\ref{fig:velocityStatistics}A), and shear deformation may be negligible (see inset (ii) of figure\ \ref{fig:velocityStatistics}B).
The model presented in section\ \ref{sec:aggregates} alternates between near-wall and bulk flows to describe stretching, utilizing the constrained near-wall flow characteristics to calculate stretching (section\ \ref{sec:isolated}).
A similar approach separating near-wall and bulk flow regions to evaluate deformation could be useful in other discrete structures, including three-dimensional media, where the coordination of near-wall shear and flow helicity density in the bulk can give rise to chaotic advection \citep{turuban_spacegroup_2018,lester_unified_2025}.
\subsection{Evaluation of the full elongation distribution}
The early studies of fluid elongation were in turbulent flows \citep{batchelor_effect_1952}, where the governing equation is $\v{z}(t) =\exp_\text{T}[\int_0^t dt' \tensor{L}(t')]\v{z}(0)$, which involves the time-ordered exponential of the integrated velocity gradient tensor \citep{cocke_turbulent_1969}.
This equation is inherently multiplicative, since the exponential splits over weakly-correlated time intervals \citep{drummond_turbulent_1990}.
Accordingly, unsteady flows readily produce log-normal-like elongation distributions, as seen in a diversity of contexts \citep{kraichnan_convection_1974, khakhar_fluid_1986, drummond_turbulent_1990, girimaji_materialelement_1990, muzzio_mixing_1992, arratia_statistics_2005, subramanian_statistics_2009, souzy_stretching_2017}.

In steady flows, the Poincaré-Bendixson theorem ensures that streamlines cannot cross, which in the Protean-frame renders the velocity gradient tensor upper triangular; the constraint of non-crossing streamlines thereby simplifies the time-ordered exponential and modifies the multiplicative structure of the stretching equations \citep{dentz_coupled_2016, lester_fluid_2018}.
In two dimensions, the equations become additive (equations\ \eqref{eq:stretch}), so there is no obvious reason to expect log-normal stretching.
In three dimensions, the equations retain multiplicative structure as the transverse stretching remains multiplicative, while the longitudinal shear accumulates as a time integral over the transverse deformation components \citep{lester_fluid_2018}.
Accordingly, three-dimensional steady flows can still produce log-normal-like stretching \citep{lester_line_2025}, as observed both in porous media \citep{souzy_velocity_2020} and static mixers \citep{hobbs_kenics_1997, hobbs_mixing_1997}.
The same conclusion of log-normal stretching can be inferred for two-dimensional unsteady flows \citep{lesterFluidDeformationRandom2026}.

The stochastic stretching model developed in section\ \ref{sec:aggregates} represents the total elongation as the sum of independent, power-law-distributed stretching increments.
According to the generalized central limit theorem, the sum of such heavy-tailed variables converges asymptotically to a Lévy-stable distribution \citep{zaburdaev_levy_2015}, as recovered by the $s\rightarrow 0$ Laplace inversion of equation\ \eqref{eq:mw}.
However, the maximum elongation of lamellae is also bounded by the observation time, creating a time-dependent truncation in the elongation increments (equation\ \eqref{eq:stretchsinglecyl}).
Truncated Lévy processes have ``ultra-slow'' convergence to the asymptotic limit, with typically three orders of magnitude slower convergence compared to the case without truncation which requires typically $\sim10$ events \citep{mantegna_stochastic_1994}, meaning times exceeding $10^3 t_a$ could be required to observe the asymptotic elongation distribution.
Consequently, a pre-asymptotic distribution likely dominates over the timescales examined here, explaining why the observed distributions lack the heavy power-law tails of the Lévy-stable form.
Furthermore, our model assumes identical elongation statistics per cylinder encounter.
This single-cylinder approximation suffices for predicting the asymptotic moments in our relatively-dilute cylinder arrays, which are largely controlled by the most-stretched lamellae in the tail of the stretching distribution \citep[cf.][]{vezzani_singlebigjump_2019}.

Arrays with touching or near-touching inclusions will produce different elongation behavior.
For pore throats much smaller than $a$, the deformation kernel $K$ still localizes near solid boundaries (Figure~\ref{fig:concept}), but the single-cylinder model becomes a crude approximation.
This breakdown may explain the continuous scaling exponent $\langle \rho \rangle \sim t^\beta$ ($0.5 \leq \beta \leq 2$) observed in random-permeability Darcy flows \citep{leborgne_lamellar_2015}.
Accounting for more structural variability in stretching increments would likely broaden the core of the elongation distribution, further altering its pre-asymptotic shape. Analyzing the pre-asymptotic elongation distribution in additional detail remains as a subject for future work.

\subsection{Implications for mixing in heterogeneous media}
Studies of transport and reactions often use randomly-arranged cylinders as prototypical porous media \citep{anna_mixing_2014, jimenez-martinez_impact_2017, willingham_evaluation_2008,meigelDispersiveTransportDynamics2022, borgman_solute_2023}.
Some of these studies have represented pore-scale elongation as simple shear flow \citep{anna_mixing_2014, jimenez-martinez_impact_2017, borgman_solute_2023}, with a shear rate set by the pore size and mean flow velocity.
In contrast to this linear-stretching assumption and in accord with \citet{dentz_coupled_2016}, we found quadratic-in-time scaling for the mean fluid elongation in disordered cylinder arrays (figure~\ref{fig:stretchingmoments}), and we described this nonlinear elongation from the accumulated encounters of the plume with low-velocity regions near cylinder walls (section~\ref{sec:aggregates}).
This quadratic stretching arises because each time the plume wraps a cylinder, two branches are generated, and each branch stretches as simple shear.
The wrap rate ($\nu_w$) grows with medium heterogeneity ($\varepsilon$), so although quadratric scaling exists for all $\varepsilon\neq0$, elongation grows with a larger pre-factor in more disordered cylinder arrays.

Nonlinear elongation in disordered media has important implications for relating the rate of mixing to the Péclet number characterizing the relative importance of advection and diffusion.
According to the lamellar theory and neglecting lamella aggregation \citep{leborgne_lamellar_2015, villermaux_mixing_2019}, for algebraic mean elongation ($\langle\rho\rangle \sim t^\alpha$), the time $t_\text{mix}$ at which concentrations begin to decay obeys $t_\text{mix}\sim\Peclet^{-2/3}$ in ordered media ($\alpha=1$) and $t_\text{mix}\sim\Peclet^{-4/5}$ in disordered media ($\alpha=2$).
Medium heterogeneity can therefore markedly speed up mixing, a characteristic not captured by the simple shear assumption.
For example, by comparing $t_\text{mix}$ between ordered and disordered media at $\Peclet\simeq 10^3$, which is a typical value in groundwater and laboratory settings \citep{bijeljic_porescale_2006}, disorder shortens the mixing time by about \SI{60}{\percent}. 
At the same time, additional elongation packs more lamellae into each pore (see inset of figure~\ref{fig:stretchingmoments}B), increasing the chance of lamella aggregation \citep{villermaux_mixing_2003, heyman_mixing_2024}.
\section{Conclusion}
Fluid stretching in two-dimensional steady flows through porous media is governed by the ratio of shear and squared velocity, a quantity that localizes near solid boundaries.
This localization enables an analytical representation of stretching through random cylinder arrays as a random alternation of lamellae into and out of near-wall regions, giving a mean elongation $\langle \rho\rangle$ proportional to $t$ for flow along symmetry directions of  ordered arrays and $t^2$ for flow through disordered arrays, regardless of the amount of medium heterogeneity.
The rapidly-growing higher moments ($\langle \rho^q \rangle$, of order $q=1,2,\ldots$) scale as $t^{3q-2}$ (ordered) and $t^{3q-1}$ (disordered). 
We have described this nonlinear and microstructure-dependent fluid stretching using a random-walk framework, showing how near-wall shear and the frequency of wall encounters determine the bulk elongation statistics.
This framework reveals that the common representation of stretching in disordered cylinder arrays as a simple shear flow is incorrect, instead showing a non-linear mean elongation growth in time consistent with the Protean-frame elongation theory developed by \citet{dentz_coupled_2016}.
The essential mechanism for quadratic elongation is that wraps over cylinders accumulate linearly in time and generate new branches, while each of these branches stretches as simple shear.

By combining the statistics of single-cylinder stretching with a medium-dependent boundary-encounter frequency, our framework also opens pathways to describe the full distribution of elongations. 
The observed elongation distributions are not log-normal over the observed timescales, consistent with the additive stretching mechanisms in two-dimensional steady flows. 
Instead, the developed theory predicts an asymptotic Lévy stable distribution, although with an extremely slow convergence due to the time-dependent truncations of the underlying elongation-increment distributions.
 
Algebraic stretching $\langle \rho\rangle \sim t^\alpha$ implies a mixing time $t_{\rm mix} \sim \Peclet^{-2\alpha/(1+2\alpha)}$ in mean-field approximation when neglecting lamella aggregation, meaning $t_{\rm mix} \sim \Peclet^{-2/3}$ for ordered media and $t_{\rm mix} \sim \Peclet^{-4/5}$ for disordered media. The efficiency of mixing markedly differs between ordered and disordered media owing to the increased number of solid-boundary encounters in disordered media. For example, at $\Peclet=10^3$ and neglecting lamella aggregation, the mixing time is $2.5\times$ shorter in disordered media. At the same time, increased stretching packs more lamellae into the pores, which will increase the propensity for lamella aggregation \citep{villermaux_mixing_2003, heyman_mixing_2024}. Whether these predicted differences in mixing from medium heterogeneity will survive lamella aggregation remains an important open prospect, as does the extension of our framework to other porous structures.

\begin{bmhead}[Acknowledgements]
KP and GL thank Tomás Aquino, Marco Dentz, Joachim Mathiesen, and Marcel Moura for valuable discussions, and the Centre for Advanced Study at the Norwegian Academy of Science and Letters for their hospitality.
\end{bmhead}
\begin{bmhead}[Funding]
This work was funded by the Research Council of Norway from grants 325819 (FRINATEK---M4), 262644 (Center of Excellence---Porelab), and 353372 (FRIPRO---MinMix). The work was also supported by the Centre for Advanced Study in Oslo, through the Young CAS project \emph{Mixing by Interfaces} to GL.
\end{bmhead}
\begin{bmhead}[Declaration of interests]
The authors report no conflict of interest.
\end{bmhead}
\begin{bmhead}[Data availability statement]
The divergence-corrected flow fields and simulation/analysis scripts used in this study are freely available at \url{https://doi.org/10.5281/zenodo.20027043}. Our software for divergence correction of porous-media flow fields \citep{pierce_deldotnot_2025} can be downloaded from \url{https://github.com/jkpierce/deldotnot}.
\end{bmhead}
\begin{bmhead}[Author ORCIDs]
K. Pierce, \url{https://orcid.org/0000-0003-3670-8636}; T. Le Borgne, \url{https://orcid.org/0000-0001-9266-9139}; F. Renard, \url{https://orcid.org/0000-0002-5125-5930}; G. Linga, \url{https://orcid.org/0000-0002-0987-8704}.
\end{bmhead}
%
%
\begin{appen}
\section{Characterization of porous structures}
\label{sec:app_throats}
The coordinates of two neighboring cylinders in the medium (labeled 0 and 1) can be written as
\begin{equation}
\begin{aligned}
    \v{x}_0 &= \frac{a\varepsilon}{2}\sqrt{r_0}(\cos\theta_0, \sin\theta_0) \\
    \v{x}_1 &= (2a,0)+\frac{a\varepsilon}{2}\sqrt{r_1}(\cos\theta_1, \sin\theta_1), 
\end{aligned}
\label{eq:perturbedcoords}
\end{equation}
where $r_0$, $r_1$, $\theta_0$, and $\theta_1$ are all independent uniform random variables, with $0\leq r_i\leq 1$ and $0\leq \theta_i\leq 2\pi$. The square root factors in equation\ \eqref{eq:perturbedcoords} ensure uniform-area sampling in the perturbation of each lattice site. The pore-throat size can be computed as $\lambda = |\v{x}_1-\v{x}_0|-a$, which to lowest order in $\varepsilon$ is
\begin{equation}
    \frac{\lambda}{a} = 1+\frac{\varepsilon c}{2},
    \label{eq:throatsize}
\end{equation}
where the random variable $c=(\sqrt{r_1}\cos\theta_1-\sqrt{r_0}\cos\theta_0)/2$ combines the original four. This low-order expansion reliably approximates the nearest-neighbor distance.
Note that $-1 \leq c \leq 1$. The distribution of $c$ can be evaluated by integrating out the four independent random variables involved in it, giving the exact result
\begin{equation}
    P(c) = \left(\frac{4}{\pi}\right)^2\int_{\text{max}(-1/2,-1/2+c)}^{\text{min}(1/2,1/2-c)}\sqrt{1-4z^2}\sqrt{1-4(z-c)^2}dz.
\end{equation}
The integral can be evaluated by steepest descent, giving the approximation:
\begin{equation}
    P(c) \approx \frac{(1-c^2)^2}{\sqrt{1+c^2}}. \label{eq:stats}
\end{equation}
This distribution can be transformed to the throat-size distribution through equation\ \eqref{eq:throatsize}. In particular, the mean throat size can be computed as $\bar{\lambda}=a$, and its relative variation can calculated as $\sigma_\lambda/\bar{\lambda} \approx 0.182 \varepsilon$, showing a linear relationship between the disorder parameter and the throat size fluctuations.
\section{Deformation kernel in the inner region}\label{sec:app_kernel}
For a particle started near a cylinder at initial coordinates $(\delta_0,\theta_0)$, the trajectories (equation\ \eqref{eq:trajs}) show two short-lived durations $(Ab_0)^{-1}$ near the stagnation points on the cylinder where the radial motion dominates ($|u_\delta|\gg |u_\theta|$). The first of these periods implies that the initial streamwise velocity is given by $u(0)=|u_r(0)|=A\delta_0^2$, independent of $\theta_0$.Otherwise, for the majority of the time the particle spends in the inner region, the angular motion dominates, and the movement direction approximately aligns with increasing $\theta$, so that $u \approx |u_\theta|$. In this regime, the velocity gradient tensor can be computed to lowest order in $\delta$ as:
\begin{equation}
\tensor{L} \approx  A\begin{pmatrix}
-2 \delta\cos \theta & -2\delta\sin \theta  \\ 
 \sin \theta & 2\delta\cos \theta
\end{pmatrix}.
\end{equation}
This tensor is already in a streamline frame to lowest order ($\tensor{L}'\approx \tensor{L}$), since the flow is primarily tangential.
Note that $\mathrm{Tr}(\tensor{L})=0$ as required by incompressibility.
The non-dimensionalized streamwise shear to lowest order is therefore $\sigma\approx A\sin\theta$, which combines with $u_0 \approx A\delta_0^2$, $\delta\sqrt{\sin\theta} = \delta_0\sqrt{ \sin \theta_0}$ (streamline constraint), and $u(t)\approx|u_\theta(t)|$ to provide the constant deformation kernel equation\ \eqref{eq:kernel1} after restoring dimensions, valid throughout the tangential-motion regime.

We can also evaluate the deformation kernel distribution by calculating $K$ without the streamline constraint, obtaining $K\sim (\delta^2\sin\theta)^{-1}$.
With uniformly-distributed $\delta$ and $\theta$, this gives $P(K)\sim K^{-3/2}$, implying that isolated cylinders and porous models have identical large-deviation statistics (see inset (i) of figure\ \ref{fig:velocityStatistics}B).
\section{Stretching statistics in cylinder arrays}
\label{sec:app_lognorm}
To calculate the asymptotic elongation moments from the Laplace-transformed distribution of equation\ \eqref{eq:mw}, we note that the large $t$ scaling of the moments relates to the $s\rightarrow 0$ scaling of the distribution via standard Tauberian theorems.
The small-$s$ inner-region densities in equation\ \eqref{eq:mw} can be calculated as:
\begin{equation}
\begin{aligned}
      \tilde{r}_I(g,s) &\simeq \tilde{\psi}_I(s) + 2\sum_{q=1}^\infty \frac{(-g\rho_\ast)^q}{q!} (s\tau)^{2-3q}\Gamma(3q-2),\\
      \tilde{R}_I(g,s) &\simeq \frac{1-\tilde{\psi}_I(s)}{s}, 
\end{aligned}
\end{equation}
where $\tilde{\psi}_I$ is the Laplace transform of equation\ \eqref{eq:innerLayerResidenceTime}.
The result for $\tilde{r}_I$ can be obtained by Taylor expanding the integrands in its Laplace transforms and integrating term by term.

The Laplace-transformed elongation moments of order $q$ derive from $\langle \tilde{\rho}(s)^q\rangle= \lim_{g\rightarrow 0}(-\partial_g)^q \tilde{P}(g,s)$.
Repeated application of $-\partial_g$ to equation\ \eqref{eq:mw} provides the recursion relation
\begin{equation}
     (-\partial_g)^q \tilde{P}(g,s) = \frac{e^{-g}[(\nu_w \!+\! t_as)\tilde{R}_I \!+\! t_a \tilde{r}_I] + \sum_{k=0}^{q-1} \binom{q}{k} [t_a e^{-g} \!+\! \nu_w (-\partial_g)^{k} \tilde{P}] (-\partial_g)^{q-k}\tilde{r}_I}{t_as+\nu_w(1-\tilde{r}_I)}.
\end{equation}
For $s\rightarrow 0$ (long times), the $k=0$ term dominates in the sum, and the Laplace-space moments become
\begin{equation}
     \langle \tilde{\rho}(s)^q\rangle \simeq \frac{1}{s} + 2\tau \rho_\ast^q\left[ (s\tau)^{1-3q} +  f_I(s\tau)^{-3q}\right] \Gamma(3q-2),
     \label{eq:laplacemoments}
\end{equation}
where we have used $\tilde{R}_I(s)\sim 2\tau$, $\tilde{r}_I(0,s)\sim1-2\tau s$, and $\tilde{P}(0,s)\sim 1/s$.
The mean residence time is $2\tau$ in inner regions and $t_a/\nu_w$ in outer regions.
We introduced the notation $f_I = 2\tau/[t_a/\nu_w+2 \tau]$ for the expected fraction of time lamellae spend in inner regions.
Because most lamellae have short residence times in the inner region, the mean time between subsequent wraps will be much larger than the average inner-region residence time $(t_a/\nu_w\gg2\tau)$, meaning $f_I \approx 2\tau\nu_w/t_a$. 
The temporal moments follow from the inverse Laplace transform of equation\ \eqref{eq:laplacemoments}:
\begin{equation}
     \langle \rho(t)^q \rangle \simeq 1 + \frac{2\rho_\ast^q}{3q-2} \left(\frac{t}{\tau}\right)^{3q-2} + \frac{4\tau\nu_w\rho_\ast^q}{(3q-1)(3q-2)t_a} \left(\frac{t}{\tau}\right)^{3q-1}.
     \label{eq:allmoments}
\end{equation}
Finally, equation\ \eqref{eq:allmoments} provides equation~\eqref{eq:collectivemeans} from $q=1$ and equation~\eqref{eq:collectivevars} from $q=2$.
Significantly, equation \eqref{eq:allmoments} applies regardless of medium heterogeneity by selecting the dominant-order term for the ordered case ($\nu_w=0$) or the disordered case ($\nu_w > 0$).
\end{appen}
\bibliographystyle{jfm}
\bibliography{biblio}
\end{document}